\documentclass[a4paper, notitlepage, 10pt]{article}
\usepackage{geometry}
\usepackage{setspace,relsize}
\fontfamily{times}
\usepackage{moreverb}

\usepackage{amsthm}
\usepackage[conf={text link section,restate}]{proof-at-the-end}
\usepackage{amsmath}
\usepackage{amssymb}
\usepackage{listings}
\usepackage{enumerate}
\usepackage{bbm}

\usepackage{indentfirst}
\usepackage{multirow}
\usepackage{url}
\usepackage{hyperref}
\hypersetup{colorlinks=true,citecolor=blue}
\usepackage{xcolor}
\usepackage{todonotes}
\usepackage{ulem}
\usepackage[authoryear,round]{natbib}

\usepackage{graphicx}
\title{\vspace{-9ex}\centering \bf Bivariate beta distribution: parameter inference and diagnostics}
\author{
Lucas Machado Moschen, Luiz Max Carvalho\\
School of Applied Mathematics, Fundação Getulio Vargas.}
\date{2 March 2023}

\newcommand{\R}{\mathbb{R}}

\newcommand{\dd}{\boldsymbol{Z}}
\newcommand{\parameter}{\boldsymbol{\alpha}}
\newcommand{\one}{\mathbbm{1}}

\newcommand{\ev}{\mathbb{E}}
\newcommand{\pr}{\mathbb{P}}
\newcommand{\var}{\operatorname{Var}}
\newcommand{\cor}{\operatorname{Cor}}
\newcommand{\cov}{\operatorname{Cov}}

\newtheorem{proposition}{Proposition}[]
\newtheorem{remark}{Remark}[]
\theoremstyle{definition}

\definecolor{keywords}{RGB}{255,0,90}
\definecolor{comments}{RGB}{0,0,113}
\definecolor{red}{RGB}{160,0,0}
\definecolor{green}{RGB}{0,150,0}

\begin{document}
\maketitle

\begin{abstract}
Correlated proportions appear in many real-world applications and present a unique challenge in terms of finding an appropriate probabilistic model due to their constrained nature.
The bivariate beta is a natural extension of the well-known beta distribution to the space of correlated quantities on ${[0, 1]}^2$.
Its construction is not unique, however.
Over the years, many bivariate beta distributions have been proposed, ranging from three to eight or more parameters, and for which the joint density and distribution moments vary in terms of mathematical tractability.
In this paper, we investigate the construction proposed by~\cite{olkin2015constructions}, which strikes a balance between parameter-richness and tractability.
We provide classical (frequentist) and Bayesian approaches to estimation in the form of method-of-moments and latent variable/data augmentation coupled with Hamiltonian Monte Carlo, respectively.
The elicitation of bivariate beta as a prior distribution is also discussed.
The development of diagnostics for checking model fit and adequacy is explored in depth with the aid of Monte Carlo experiments under both well-specified and misspecified data-generating settings.

Keywords: Bivariate beta; correlated proportions; Diagnostics; Method of moments; Bayesian estimation. 
\end{abstract}

\section{Introduction}

Correlated proportions appear in many real-world applications such as modelling pollen distributions in forests~\citep{nadarajah_new_2017}, the relationship between drought frequency and duration~\citep{nadarajah2007new} and sensitivity/specificity of imperfect detection systems (e.g.\ disease tests,~\cite{dahabreh2013empirical}).
The beta distribution is a widely used uni-dimensional distribution for random variables with support over $[0,1]$, and extensions to the square $[0,1] \times [0,1]$  are natural, yielding the bivariate beta distributions, {\it i.e.}, bivariate distributions with beta-distributed marginals.
Many constructions are possible and vary depending on the number of needed parameters, the attainable correlation structure, and the mathematical and computational tractability of the joint probability density function (pdf, see below).

% \subsection{Related work}

~\cite{balakrishnan2009continuous} present an extensive review of continuous bivariate distributions, in special those with beta marginals, from the Dirichlet distribution to the construction through copulas.
~\cite{trick2021bivariate} provide an updated assessment on the topic with a focus on those built through transformations of gamma-distributed random variables.
An incomplete list of papers dealing with bivariate beta distributions is~\cite{libby1982multivariate},~\cite{olkin2003bivariate},~\cite{magnussen2004algorithm},~\cite{nadarajah2005some},~\cite{sarabia2006bivariate},~\cite{nadarajah2007new},~\cite{arnold2011flexible},~\cite{nadarajah_new_2017} and~\cite{trick2021bivariate}.

A straightforward and popular way of introducing a correlation between beta marginals is using shared gamma-distributed random variables. 
~\cite{libby1982multivariate} discussed a multivariate distribution where the marginals are generalised beta distributions with three parameters, the joint density is available in closed form and the moments are represented through series. 
For the bivariate case, this is a six-parameter distribution.
~\cite{olkin2003bivariate} introduced a particular case of this distribution using only three parameters.
~\cite{sarabia2006bivariate} extended this result by studying several bivariate distributions with generalised beta-distributed marginals.
They argued that a three-parameter distribution is necessary to model the mean, the variance and the skewness of the marginals.
~\cite{nadarajah2007new} constructed a different bivariate beta distribution with generalised beta distributions as marginals, for which the moments have no closed-form expression and variate simulation necessitates rejection sampling.
On the other hand, the joint density, up to a constant that depends on the parameters, has an analytical expression.
~\cite{nadarajah2007new} then proposed a maximum likelihood and method of moments estimators for their bivariate beta construction.

~\cite{arnold2011flexible} proposed a five-parameter model from independent gamma distributions with a common scale parameter. 
One desirable aspect of this distribution is the possibility of the correlation between the margins assuming the full range $(-1,1)$, which sets it apart from the previous constructions.
However, the density and the product moments are not available in closed form.
Because of that,~\cite{arnold2011flexible} applied a modified maximum likelihood estimator, in which marginal parameters are estimated by maximising the marginal likelihood and the method of moments is used for estimating the correlation parameter.
This distribution presents two problems: (i) parameter estimates can be negative for the method of moments, and a heuristic solution is to set them to be $0$ when this happens; (ii) it does not allow for arbitrary beta marginals.
The authors employ a Monte Carlo simulation study to evaluate all the variations in the estimation process.
Another problem with~\cite{arnold2011flexible} --- discussed by~\cite{olkin2015constructions} --- is the difficulty to extend the distribution to higher dimensions.  

~\cite{nadarajah_new_2017} presented a six-parameter distribution with elementary pdfs and argued numerically that the correlation coefficient covers the range $[0,1)$, despite not presenting a mathematical proof.
The product moments have a double infinite sum representation and the marginals are a generalisation of the beta distribution.
The parameter estimation is through maximum likelihood.
In particular, the authors compared it with other bivariate beta distributions with elementary pdfs, such as~\cite{libby1982multivariate},~\cite{sarabia2006bivariate} and~\cite{nadarajah2007new}. 
Recently,~\cite{trick2021bivariate} studied a six-parameter bivariate beta extending the work from~\cite{magnussen2004algorithm}.
This distribution models arbitrary beta marginals with the drawback of allowing only positive correlations and the product moments not being available in a closed form.
The authors remark that two different parameter specifications generate similar data, which leads to an identifiability problem in practical settings.
As a solution for parameter inference they reduced the distribution to five parameters, imposing a constraint on the parameter space.    

~\cite{barros2015estimaccao} proposed an estimation method for the bivariate beta presented in~\cite{nadarajah2005some}, which is based on the fact that the product of independent beta-distributed random variables is also beta-distributed.
In this formulation, the moments are written in closed form, but the joint  pdf is not. 
Moreover, the correlation is strictly positive in this case.
~\cite{crackel2017bayesian} proposed a Bayesian approach for the estimation of the parameters from~\cite{arnold2011flexible}'s distribution and an extension to the original model with eight parameters.
Since the likelihood is intractable, they use an approximate Bayesian computation (ABC) algorithm to approximate the posterior distribution.

The presentation so far has made it clear that one needs to strike a balance between parameter-richness and tractability: some constructions have many parameters and are flexible, allowing for a full range of correlations, for instance.
In many cases, however, the joint density is not tractable, complicating likelihood-based methods.
In other instances, the moments are not closed-form thus impeding efficient method-of-moments techniques and straightforward interpretations of the impact of the parameters on the moments.
In what follows we will detail the construction of~\cite{olkin2015constructions}, which we argue achieves a good balance between tractability and parameter-richness.

~\cite{olkin2015constructions} describe a bivariate distribution with beta-distributed marginals, positive probability over the space $(0,1) \times (0,1)$, and correlation over the full range $(-1,1)$.
Its construction is the following: let $\boldsymbol{U} \sim \operatorname{Dirichlet}(\boldsymbol{\alpha})$, with $\boldsymbol{U} = (U_1, U_2, U_3, U_4)$ and $\boldsymbol{\alpha} = (\alpha_1, \alpha_2, \alpha_3, \alpha_4)$, such that $\alpha_i > 0$ for $i = 1,\dots,4$ and $U_4 = 1 - U_1 - U_2 - U_3$. 
The joint density of $\boldsymbol{U}$ with respect to the Lebesgue measure on $\R^3$ is given by
\begin{equation}
  \label{eq:dirichlet-distribution}
  f_U(u_1, u_2, u_3) = \frac{1}{B(\boldsymbol{\alpha})}u_1^{\alpha_1-1}u_2^{\alpha_2-1}u_3^{\alpha_3-1}{(1-u_1-u_2-u_3)}^{\alpha_4-1} , 
\end{equation}
when $u_i \in [0,1]$ for $i = 1,2,3$ and $u_1 + u_2 + u_3 \le 1$.
Otherwise, $f_U(u_1, u_2, u_3) = 0$.
The normalising constant is defined as
\[
B(\boldsymbol{v}) = \frac{\prod_{i=1}^n \Gamma(v_i)}{\Gamma\left(\sum_{i=1}^n v_i\right)},
\]
for $\boldsymbol{v} \in \R^n$ with positive entries.

Let $X = U_1 + U_2$ and $Y = U_1 + U_3$.
In~\autoref{sec:theory}, we show that the marginal distributions of $X$ and $Y$ are betas and thus the bivariate random vector $(X, Y)$ has a bivariate beta distribution with parameter $\boldsymbol{\alpha}$.
This is a four-parameter construction with a full range of correlations between the margins and for which the joint density is intractable but the moments have closed-form expressions.
Moreover, random variates are easy to simulate, allowing for easy implementation of inferential tasks such as prediction.

In this paper, we therefore centre our attention on the parameter estimation for bivariate beta distribution proposed in~\cite{olkin2015constructions}, following frequentist (method of moments) and Bayesian approaches (latent variable representation coupled with Hamiltonian Monte Carlo).
We also address an important but often neglected aspect of inference which is the development of \textbf{diagnostics} for checking model fit and adequacy.

\subsection{Contributions}\label{sec:contributions}

In this paper, we provide a deep investigation of parameter estimation for the distribution introduced by~\cite{olkin2015constructions}.
In particular, we study parameter estimation when a $n$-sized random sample $\dd = \{ (X_1, Y_1), \ldots, (X_n, Y_n) \}$ is available and one would like to estimate $\boldsymbol{\alpha}$.
% Moreover, we use this discussion for the problem of elicitation in the Bayesian setting.
Our main contributions are three-fold:
\begin{itemize}
    \item Exact solution for the method of moments estimator when it exists, identification of the space where the moments do not yield a well-defined bivariate beta and four classes of moment-based estimators;
    
    \item Bayesian estimation approximating the posterior distribution $p(\boldsymbol{\alpha} \mid \dd)$ through Hamiltonian Monte Carlo in the Stan programming language~\citep{carpenter2017stan,stan}. 
    We propose a latent-variable representation which circumvents the intractability of the likelihood and leads to an efficient exploration of the posterior.
    
    \item Custom diagnostics to identify model fit problems under both the classical (frequentist) and Bayesian approaches. 
\end{itemize}

\section{Theory}\label{sec:theory}

Let $(X, Y)$ have the bivariate beta distribution as constructed by~\cite{olkin2015constructions},  with parameter $\parameter$.
By the aggregation property~\cite[Section 2.2]{ng2011dirichlet},
\[
(U_1 + U_2, U_3, U_4) \sim \operatorname{Dirichlet}(\alpha_1 + \alpha_2, \alpha_3, \alpha_4),
\]
which has beta-distributed marginals, implying that $X \sim \operatorname{Beta}(\alpha_1 + \alpha_2, \alpha_3 + \alpha_4)$.
Similarly, $Y \sim \operatorname{Beta}(\alpha_1 + \alpha_3, \alpha_2 + \alpha_4)$, which establishes that the marginal distributions are beta-distributed.
From this construction, we can compute the means, the variances and other moments of $X$ and $Y$.
Denoting $s_{\alpha} = \sum_{i=1}^4 \alpha_i$, we have 
\begin{gather}
    \begin{aligned}
    \ev[X] &= \frac{\alpha_1 + \alpha_2}{s_{\alpha}},
    & \ev[Y] &= \frac{\alpha_1 + \alpha_3}{s_{\alpha}},
    \\
    \var(X) &= \frac{(\alpha_1 + \alpha_2)(\alpha_3 + \alpha_4)}{s_{\alpha}^2(s_{\alpha} + 1)},
    & \var(Y) &= \frac{(\alpha_1 + \alpha_3)(\alpha_2 + \alpha_4)}{s_{\alpha}^2(s_{\alpha} + 1)}.
    \end{aligned}
\end{gather} 

The sum of the parameters of the marginal distributions is the same for $X$ and $Y$, which restricts the family of pairs of beta distributions for the marginals that generate a well-defined bivariate beta distribution.
The density of $(X, Y)$ is given in~\autoref{prop:bivariate-beta-density} and, as far as we know, does not have a closed-form expression.
It can, however, be expressed with special functions --- see Appendix A of~\cite{olkin2015constructions}.
\autoref{fig:beta-bivariate} illustrates the density for different values of $\boldsymbol{\alpha}$.

\begin{theoremEnd}{proposition}[Bivariate beta density]\label{prop:bivariate-beta-density}
  The joint density of $(X, Y)$ with respect to the Lebesgue measure on $\R^2$ is given by 
  \begin{equation}
    \label{eq:dist-X-Y}
    f_{X,Y}(x,y) = \frac{1}{B(\boldsymbol{\alpha})}\int_{\Omega} u^{\alpha_1 - 1}{(x - u)}^{\alpha_2 -1}{(y-u)}^{\alpha_3-1}{(1-x-y+u)}^{\alpha_4-1} \, du,
  \end{equation}
  where 
  \[
  \Omega = (\max(0, x+y-1), \min(x,y)).
  \]
\end{theoremEnd}

\begin{proofEnd}
   Note that
   \[
  \begin{bmatrix}
    U_1 \\ X \\ Y
  \end{bmatrix}  = \begin{bmatrix}
    1 & 0 & 0 \\
    1 & 1 & 0 \\
    1 & 0 & 1
  \end{bmatrix}\begin{bmatrix}
    U_1 \\ U_2 \\ U_3
  \end{bmatrix}.
  \]
  By the Change of Variables formula, 
  \begin{equation}
    \begin{split}
      f_{U_1,X,Y}(u_1,x,y) &= f_{U_1,U_2,U_3}(u_1, x - u_1, y - u_1), \\ 
      &= \frac{1}{B(\boldsymbol{\alpha})}u_1^{\alpha_1-1}{(x-u_1)}^{\alpha_2-1}{(y-u_1)}^{\alpha_3-1}{(1-x-y+u_1)}^{\alpha_4-1},
    \end{split}
  \end{equation}
  where $0 \le u_1 \le x, u_1 \le y$, and $0 \le 1 - x - y + u_1$.  
  Hence,
  \begin{equation}
      f_{X,Y}(x,y) = \frac{1}{B(\boldsymbol{\alpha})}\int_{\Omega} u_1^{\alpha_1-1}{(x-u_1)}^{\alpha_2-1}{(y-u_1)}^{\alpha_3-1}{(1-x-y+u_1)}^{\alpha_4-1} \, du_1,
  \end{equation}
  such that $\Omega = \{u_1 \in [0,1] : \max(0, x + y -1) < u_1 < \min(x,y)\}$.
\end{proofEnd} 

As noted in Section 2.3 of~\cite{olkin2015constructions}, the non-central moments can be calculated through the formula
\[
\ev[X^r Y^s] = \ev[{(U_1 + U_2)}^r{(U_1 + U_3)}^s].
\]
Combining~\cite[p. 39]{ng2011dirichlet}
\begin{equation}
    \label{eq:covariance-dirichlet}
    \cov(U_i, U_j) = -\frac{\alpha_i \alpha_j}{s_{\alpha}^2(s_{\alpha} + 1)}, i,j = 1, \dots, 4, i \neq j,
\end{equation}
with the marginal moments, we derive the following proposition:

\begin{proposition}[Covariance and correlation]\label{prop:covariance-correlation}
  The covariance between $X$ and $Y$ is given by
  \begin{equation}
      \label{eq:covariance-bivariate-beta}
      \cov(X,Y) = \frac{1}{s_{\alpha}^2(s_{\alpha}+1)}(\alpha_1\alpha_4 - \alpha_2\alpha_3),
  \end{equation}
  and the correlation is
  \begin{equation}
     \label{eq:correlation-bivariate-beta}
     \cor(X,Y) = \frac{\alpha_1\alpha_4 - \alpha_2\alpha_3}{\sqrt{(\alpha_1+\alpha_2)(\alpha_3+\alpha_4)(\alpha_1+\alpha_3)(\alpha_2+\alpha_4)}}.
  \end{equation}
\end{proposition}

In the following, we show that expression~\eqref{eq:correlation-bivariate-beta} allows the correlation between $X$ and $Y$ to span the full range $(-1,1)$. 
Fix $\rho \in (-1,1)$ and let $X$ and $Y$ be marginally distributed as $\operatorname{Beta}(a,a), a>0$. 
To have a bivariate beta distribution with these marginals, we need
\[
\alpha_1 +
\alpha_2 = \alpha_3 + \alpha_4 = \alpha_1 + \alpha_3 = \alpha_2 + \alpha_4 = a,
\]
a solution for which is $\alpha_1 = \alpha_4 \in (0,a)$ and $\alpha_2 = \alpha_3 = a - \alpha_4$. 
The correlation formula~\eqref{eq:correlation-bivariate-beta} reduces to 
\[
\cor(X,Y) = \frac{\alpha_4^2 - {(a - \alpha_4)}^2}{a^2} = \frac{2}{a}\alpha_4 - 1,
\]
which implies that $\alpha_4 = a\dfrac{1 + \rho}{2} \in (0,a)$ is a solution to $\cor(X,Y) = \rho$. 
Therefore the bivariate beta with parameter $\boldsymbol{\alpha} = \frac{a}{2}(1 + \rho, 1 - \rho, 1 - \rho, 1 + \rho)$ has correlation $\rho$.

Theorem 2.1 in~\cite{ng2011dirichlet} (pp. 40) relates the Dirichlet and Gamma distributions, resulting in~\autoref{prop:relation-gamma}. 
This representation shows that this model is a limit case of the eight-parameter bivariate beta defined in Section 6.1 of~\cite{arnold2011flexible} when $\delta_1, \dots, \delta_4 \to 0$.

\begin{proposition}[Relation to the Gamma distribution]\label{prop:relation-gamma}
    Let $Y_i \sim \operatorname{Gamma}(\alpha_i, 1)$ for $i=1, \dots, 4$ and define
    \[
    X = \frac{Y_1 + Y_2}{Y_1 + Y_2 + Y_3 + Y_4} \quad \text{ and } \quad Y = \frac{Y_1 + Y_3}{Y_1 + Y_2 + Y_3 + Y_4}.
    \]
    Then $(X,Y)$ has a bivariate beta distribution with parameter $\boldsymbol{\alpha} = (\alpha_1, \dots, \alpha_4)$.
\end{proposition}

\begin{remark}\label{remark:inequality-beta}
    Suppose that $X \in (0,1)$. 
    Since $1-X > 0$, we have $\ev[X] > \ev[X^2]$, which implies that 
    \[
    \var(X) = \ev[X^2] - {\ev[X]}^2 < \ev[X] - {\ev[X]}^2 = \ev[X](1 - \ev[X]).
    \]
    Therefore, the variance of $X$ is limited by a function of the mean.
    In particular, if a variable has a beta distribution, it respects the above inequality.
\end{remark}

\subsection{Moments}\label{sec:moments}

In this section, we discuss issues related to the moments of the bivariate beta, which are important in the treatment of estimation and elicitation in Sections~\ref{sec:estimation} and~\ref{sec:elicitation}. 
Denote $m_1 = \ev[X]$, $m_2 = \ev[Y]$, $v_1 = \var(X)$, $v_2 = \var(Y)$ and $\rho = \cor(X,Y)$.
If $(X, Y)$ follows the bivariate beta distribution with parameter $\boldsymbol{\alpha}$, these quantities satisfy the following non-linear system of equations, which we call the moments' system:
\begin{equation}
  \label{eq:system-moments-alpha}
  \begin{cases}
    m_1 = \dfrac{\alpha_1+\alpha_2}{s_{\alpha}}, \\[10pt]
    m_2 = \dfrac{\alpha_1+\alpha_3}{s_{\alpha}}, \\[10pt]
    v_1 = \dfrac{(\alpha_1+\alpha_2)(\alpha_3+\alpha_4)}{s_{\alpha}^2(s_{\alpha}+1)}, \\[10pt]
    v_2 = \dfrac{(\alpha_1+\alpha_3)(\alpha_2+\alpha_4)}{s_{\alpha}^2(s_{\alpha}+1)}, \\[10pt]
    \rho = \dfrac{\alpha_1\alpha_4 - \alpha_2\alpha_3}{\sqrt{(\alpha_1+\alpha_2)(\alpha_3+\alpha_4)(\alpha_1+\alpha_3)(\alpha_2+\alpha_4)}}.\\
  \end{cases}
\end{equation}

The following proposition clarifies when the moments' system in~\eqref{eq:system-moments-alpha} has a solution for $\boldsymbol{\alpha}$, implying the existence of a bivariate beta distribution with predetermined moments. 
In particular, it delineates when we cannot define a bivariate beta following~\citeauthor{olkin2015constructions}'s approach from the moments, which is an important step for model identification.

\begin{theoremEnd}{proposition}[Solution to the moments' system]\label{prop:solution-to-system-bivariate-beta}
    System~\eqref{eq:system-moments-alpha} has no solution if
    \begin{equation}
        \label{eq:inequality-bivariate-beta}
        \frac{m_1(1-m_1)}{v_1} \neq \frac{m_2(1-m_2)}{v_2}.
    \end{equation}
    Moreover, removing the fourth equation with respect to $v_2$, the system has a unique solution given by 
    \begin{gather}\label{eq:system-solution}
    \begin{aligned}
    \alpha_1 &= (m_1 + m_2 - 1)\bar{\alpha} + \alpha_4, \\
    \alpha_2 &=  (1 - m_2)\bar{\alpha} - \alpha_4, \\
    \alpha_3 &= (1-m_1)\bar{\alpha} - \alpha_4, \\
    \alpha_4 &= \bar{\alpha}\left(\rho\sqrt{m_1m_2(1-m_1)(1-m_2)} + (1-m_1)(1-m_2)\right),
    \end{aligned}
    \end{gather}
    where
    \begin{equation}
        \label{eq:sum-alpha-solution}
        \bar{\alpha} = \frac{m_1 - m_1^2 - v_1}{v_1} = s_{\alpha}.
    \end{equation}
\end{theoremEnd}

\begin{proofEnd}
    Notice that we can simplify the third and fourth equations since 
    \[
    \frac{\alpha_3 + \alpha_4}{s_{\alpha}} = \frac{s_{\alpha} - (\alpha_1 + \alpha_2)}{s_{\alpha}} = 1 - m_1, 
    \]
    and analogously, 
    \[
    \frac{\alpha_2 + \alpha_4}{s_{\alpha}} = 1 - m_2. 
    \]
    Therefore, 
    \begin{align*}
        v_1 &= \frac{m_1(1 - m_1)}{s_{\alpha} + 1}, \\
        v_2 &= \frac{m_2(1 - m_2)}{s_{\alpha} + 1}.
    \end{align*}
    This already tells us that the system does not have a solution if 
    \[
    \frac{m_1(1-m_1)}{v_1} \neq \frac{m_2(1-m_2)}{v_2}. 
    \]
    
    The first two equations of system~\eqref{eq:system-moments-alpha} can be rewritten as a linear system:
    \begin{align*}
      (m_1 - 1)\alpha_1 + (m_1 - 1)\alpha_2 + m_1\alpha_3 + m_1\alpha_4 &= 0, \\
      (m_2 - 1)\alpha_1 + m_2\alpha_2 + (m_2-1)\alpha_3 + m_2\alpha_4 &= 0,
    \end{align*}
    which is equivalent to 
    \begin{align*}
      \alpha_1 + \alpha_2 + \frac{m_1}{m_1-1}\alpha_3 + \frac{m_1}{m_1-1}\alpha_4 &= 0, \\
      \alpha_2 + \frac{1-m_2}{m_1-1}\alpha_3 + \frac{m_1-m_2}{m_1-1}\alpha_4 &= 0.
    \end{align*}
    Then, we can write $\alpha_1$ and $\alpha_2$ as functions of $\alpha_3$ and $\alpha_4$:
    \begin{align*}
      \alpha_1 &= \frac{m_1+m_2-1}{1-m_1}\alpha_3 + \frac{m_2}{1-m_1}\alpha_4 \\
      \alpha_2 &= \frac{1-m_2}{1-m_1}\alpha_3 + \frac{m_1-m_2}{1-m_1}\alpha_4.
    \end{align*}

    Based on the expression above, denote $\alpha_1 = a_3\alpha_3 + a_4\alpha_4$, $\alpha_2 = b_3\alpha_3 + b_4\alpha_4$, $c_3 = a_3 + b_3 + 1$, and $c_4 = a_4 + b_4 + 1$. 
    Then, the third equation can be rewritten as 
    \[
    s_{\alpha} = a_3\alpha_3 + a_4\alpha_4 + b_3\alpha_3 + b_4\alpha_4 + \alpha_3 + \alpha_4 = c_3\alpha_3 + c_4 \alpha_4 = \frac{m_1(1-m_1)}{v_1} - 1, 
    \]
    which implies that 
    \[
    \alpha_3 = \frac{m_1(1-m_1) - v_1 - c_4v_1\alpha_4}{c_3v_1},
    \]
    which is a linear function of $\alpha_4$. 
    We summarise the expressions in the function of $\alpha_4$ with some simplifications: 
    \begin{align*}
      \alpha_1 &= (m_1 + m_2 - 1)\frac{(m_1 - m_1^2 - v_1)}{v_1} + \alpha_4, \\
      \alpha_2 &=  (1 - m_2)\frac{(m_1 - m_1^2 - v_1)}{v_1} - \alpha_4, \\
      \alpha_3 &= (1-m_1)\frac{(m_1 - m_1^2 - v_1)}{v_1} - \alpha_4,
    \end{align*}
    and then,
    \[
    s_{\alpha} = \frac{m_1 - m_1^2 - v_1}{v_1}.
    \]

    Now rewrite the fifth equation using the first two equations from system~\eqref{eq:system-moments-alpha} and the equations above as follows 
    \begin{equation}
        \label{eq:rho-equation}
        \begin{split}
            \rho &= \frac{\alpha_1\alpha_4 - \alpha_2\alpha_3}{\sqrt{(\alpha_1+\alpha_2)(\alpha_3+\alpha_4)(\alpha_1+\alpha_3)(\alpha_2+\alpha_4)}}, \\ 
            &= \frac{\alpha_1\alpha_4 - \alpha_2\alpha_3}{s_{\alpha}^2\sqrt{m_1m_2(1-m_1)(1-m_2)}}, \\ 
            &= \frac{(m_1 + m_2 - 1)s_{\alpha}\alpha_4 + \alpha_4^2 - ((1-m_2)s_{\alpha} - \alpha_4)((1-m_1)s_{\alpha} - \alpha_4)}{s_{\alpha}^2\sqrt{m_1m_2(1-m_1)(1-m_2)}}, \\
            &= \frac{\alpha_4 - (1-m_1)(1-m_2)s_{\alpha}}{s_{\alpha}\sqrt{m_1m_2(1-m_1)(1-m_2)}}, \\ 
        \end{split}
    \end{equation}
    and the solution is, therefore, 
    \[
    \alpha_4 = s_{\alpha}\left(\rho\sqrt{m_1m_2(1-m_1)(1-m_2)} + (1-m_1)(1-m_2)\right),
    \]
    which concludes the proof.
\end{proofEnd}
Notice that the first two equations of the system in~\eqref{eq:system-moments-alpha} imply the relations
\begin{align}
  \label{eq:alpha1-as-function-alpha3-alpha4}
  \alpha_1 &= \frac{m_1+m_2-1}{1-m_1}\alpha_3 + \frac{m_2}{1-m_1}\alpha_4 \\
  \label{eq:alpha2-as-function-alpha3-alpha4}
  \alpha_2 &= \frac{1-m_2}{1-m_1}\alpha_3 + \frac{m_1-m_2}{1-m_1}\alpha_4,
\end{align}
which are used throughout the text.

\begin{figure}[!htbp]
    \centering
    \includegraphics[width=\textwidth]{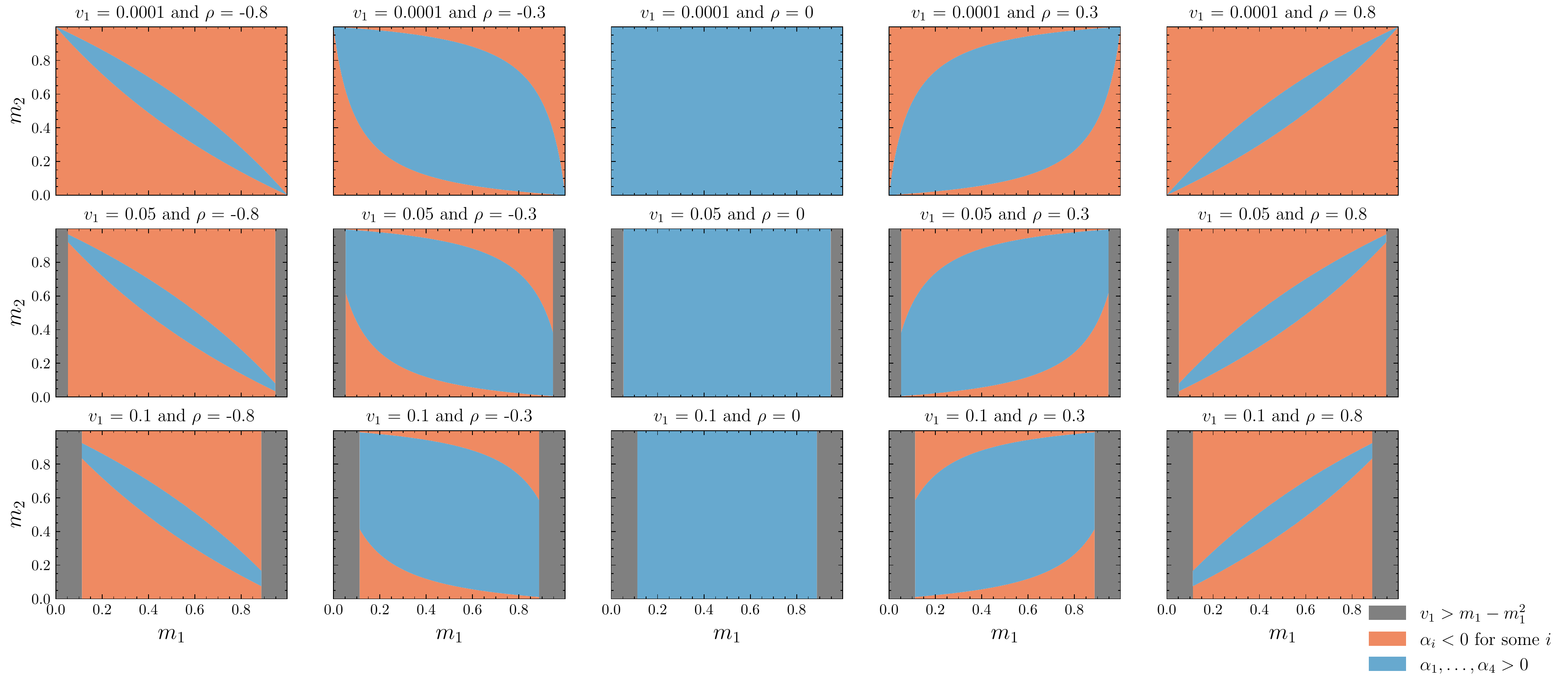}
    \caption{{\bf Positive solution for the moments' system:} 
    Representation of the sets $R_{v_1, \rho} = \{(m_1, m_2) \,:\, (m_1, m_2, v_1, \rho) \in \mathcal{M} \}$ in blue for different values of $v_1$ and $\rho$, such that $\mathcal{M}$ is the set of values $(m_1, m_2, v_1, \rho)$ which result in a well-defined bivariate beta distribution.
    In orange, we denote the regions where some $\alpha_i$ is negative in solution~\eqref{eq:system-solution}. 
    In grey, we highlight the region where the variance and the mean do not satisfy the relation from~\autoref{remark:inequality-beta}.
    }\label{fig:alpha-solutions}
\end{figure}

Besides solving the system in~\eqref{eq:system-moments-alpha}, the parameter space of the bivariate beta demands that the solution satisfies $\alpha_1, \dots, \alpha_4 > 0$.
However, this is not always achievable.
Let $\mathcal{M} \subseteq {[0,1]}^3 \times [-1,1]$ be the set of values $(m_1, m_2, v_1, \rho)$, for which the solution in~\eqref{eq:system-solution} is strictly positive \textbf{and} $v_1 < m_1(1 - m_1)$ --- see~\autoref{remark:inequality-beta}.
\autoref{fig:alpha-solutions} illustrates regions of the form
\[
R_{v_1, \rho} = \{(m_1, m_2) \,:\, (m_1, m_2, v_1, \rho) \in \mathcal{M} \},
\]
where $v_1$ and $\rho$ are fixed for each subplot.

Following~\autoref{remark:inequality-beta}, we might fix $v_1 < m_1(1-m_1)$, which implies $\bar{\alpha} > 0$.
Dividing the solution in~\eqref{eq:system-solution} by $\bar{\alpha}$, we see that $\alpha_1, \dots, \alpha_4 > 0$ if and only if (iff) the four inequalities below are satisfied:
\begin{align*}
    m_1m_2 + \rho\sqrt{m_1m_2(1-m_1)(1-m_2)} > 0 &\iff \rho > -\dfrac{m_1m_2}{\sqrt{m_1m_2(1-m_1)(1-m_2)}},\\[1ex]
    m_1(1 - m_2) - \rho\sqrt{m_1m_2(1-m_1)(1-m_2)} > 0 &\iff \rho < \dfrac{m_1(1-m_2)}{\sqrt{m_1m_2(1-m_1)(1-m_2)}},\\[1ex]
    (1- m_1)m_2 - \rho\sqrt{m_1m_2(1-m_1)(1-m_2)} > 0 &\iff \rho < \dfrac{(1-m_1)m_2}{\sqrt{m_1m_2(1-m_1)(1-m_2)}},\\[1ex]
    (1-m_1)(1-m_2) + \rho\sqrt{m_1m_2(1-m_1)(1-m_2)} > 0 &\iff \rho > -\dfrac{(1-m_1)(1-m_2)}{\sqrt{m_1m_2(1-m_1)(1-m_2)}},
\end{align*}
which yields the following interval for $\rho$:
\begin{equation}
    \label{eq:interval-rho}
    \rho \in \left(-\frac{\min(m_1m_2, (1-m_1)(1-m_2))}{\sqrt{m_1m_2(1-m_1)(1-m_2)}}, 
    \frac{\min(m_1, m_2) - m_1m_2}{\sqrt{m_1m_2(1-m_1)(1-m_2)}}\right).     
\end{equation}

\begin{figure}[!htbp]
    \centering
    \includegraphics[width=0.6\textwidth]{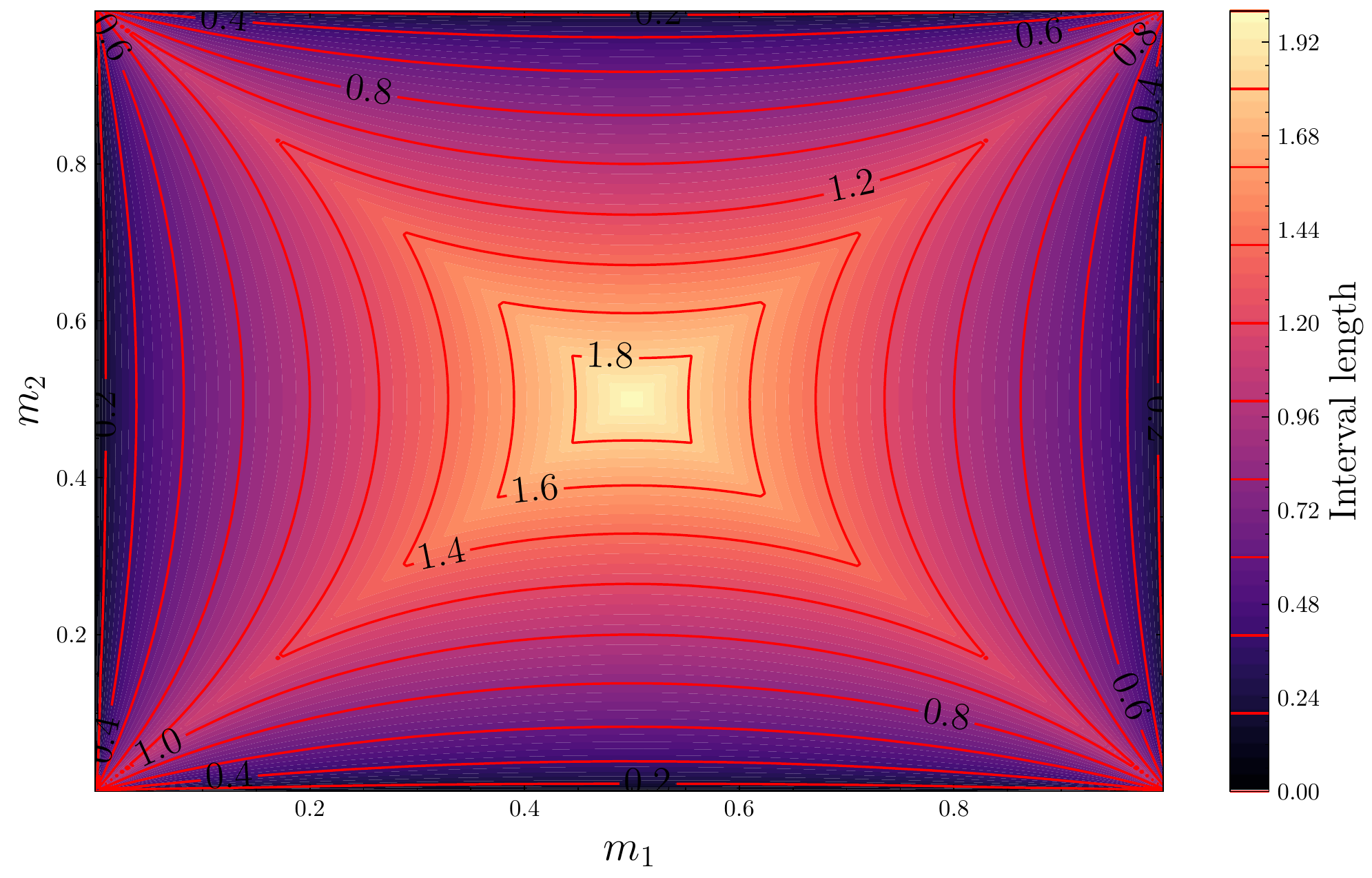}
    \caption{{\bf Interval length for $\rho$:} contour plot of the interval length for different values of $m_1$ and $m_2$ considering the expression in~\eqref{eq:interval-rho}.
    The nearer the length is to $2$ the better, indicating that we have a wider range to choose $\rho$ from given the values of the means in order to have a well-defined bivariate beta distribution. 
    }\label{fig:rho_interval_length}
\end{figure}

For instance, when $m_1 = m_2 = m$, the upper bound is $\rho_{\max} = 1$ and the lower bound is 
\[
\rho_{\min} = -\frac{\min(m, 1-m)}{\max(m, 1-m)}.
\]
On the other hand, if $m_1 = 1 - m_2 = m$, the lower bound is $\rho_{\min} = -1$ an the upper bound is 
\[
\rho_{\max} = \frac{\min(m, 1-m)}{\max(m, 1-m)},
\]
which agrees with~\autoref{fig:alpha-solutions}.~\autoref{fig:rho_interval_length} shows the interval length for different values of $m_1$ and $m_2$.
The farther $m_1$ and $m_2$ are from the centre $(0.5, 0.5)$, the smaller the interval, revealing the strong relationship between the means and the correlation coefficient.

A slight modification of~\autoref{prop:solution-to-system-bivariate-beta} solves the moments' system without the equations related to $v_1$ and $v_2$, culminating in the following proposition:

\begin{proposition}[System of three moments]\label{prop:system-three-moments}
    Considering the system in~\eqref{eq:system-moments-alpha} without the equations of $v_1$ and $v_2$, the solution is 
    \begin{equation}
      \label{eq:system-three-solution}
      \begin{aligned}
        \alpha_1 &= \alpha_4\frac{m_1m_2 + \rho\sqrt{m_1m_2(1-m_1)(1-m_2)}}{(1-m_1)(1-m_2) + \rho\sqrt{m_1m_2(1-m_1)(1-m_2)}}, \\
        \alpha_2 &= \alpha_4\frac{m_1(1-m_2) - \rho\sqrt{m_1m_2(1-m_1)(1-m_2)}}{(1-m_1)(1-m_2) + \rho\sqrt{m_1m_2(1-m_1)(1-m_2)}}, \\
        \alpha_3 &= \alpha_4\frac{m_2(1-m_1) - \rho\sqrt{m_1m_2(1-m_1)(1-m_2)}}{(1-m_1)(1-m_2) + \rho\sqrt{m_1m_2(1-m_1)(1-m_2)}},
      \end{aligned}
    \end{equation}
    with $\alpha_4$ being a free parameter.
\end{proposition}

The proof of this proposition uses relations~\eqref{eq:alpha1-as-function-alpha3-alpha4} and~\eqref{eq:alpha2-as-function-alpha3-alpha4}, and the symbolic solver SymPy~\cite[]{sympy}, as coded in~\autoref{sec:code-solution-three-equations}.
Moreover, we checked this and all the other solutions in the paper numerically.

\section{Parameter estimation}\label{sec:estimation}

We now move on to develop strategies to estimate the parameter vector $\boldsymbol{\alpha}$ of the bivariate beta distributions from data $\dd = \{ (X_1, Y_1), \dots, (X_n, Y_n) \}$, where $\dd \overset{iid}{\sim} \operatorname{Bivariate Beta}(\boldsymbol{\alpha})$. 
We will explore a method of moments and a Bayesian approach.

\subsection{Method of moments}

We start with a frequentist approach based on the method of moments by leveraging the results presented in~\autoref{sec:moments}.

Denote 
\begin{equation}
    \label{eq:empirical-moments}
    \begin{gathered}
    \bar{X}_n = \frac{1}{n}\sum_{i=1}^n X_i, \quad \bar{Y}_n = \frac{1}{n}\sum_{i=1}^n Y_i,  \quad P_n = \dfrac{\sum_{i=1}^n (X_i - \bar{X}_n)(Y_i - \bar{Y}_n)}{(n-1)S_{X,n} S_{Y,n}} \\ 
    S_{X,n}^2 = \frac{1}{n-1}\sum_{i=1}^n {(X_i - \bar{X}_n)}^2 \quad  \text{ and }\quad S_{Y,n}^2 = \frac{1}{n-1}\sum_{i=1}^n {(Y_i - \bar{Y}_n)}^2.
    \end{gathered}
\end{equation}

For the observed (sample) versions of these statistics, we write $\bar{X}_n = \hat{m}_1$, $\bar{Y}_n = \hat{m}_2$, $S^2_{X,n} = \hat{v}_1$, $S^2_{Y,n} = \hat{v}_2$ and $P_n = \hat\rho$.
The method of moments will thus proceed by matching the theoretical to the observed moments to find a solution $\boldsymbol{\alpha}$.
Therefore, we substitute $(\bar{X}_n, \bar{Y}_n, S_{X,n}^2, S_{Y,n}^2, P_n)$ for  $(m_1, m_2, v_1, v_2, \rho)$ in the equations of~\autoref{sec:moments}.
To quantify the uncertainty in our estimates by estimating standard errors and constructing approximate 95\% confidence intervals for each component of $\boldsymbol\alpha$, we employ a simple non-parametric bootstrap~\cite[]{efron1979bootstrap} method with $B=500$ re-samplings over the pairs of observations. 

\subsubsection*{Method of moments with the analytical solution for four equations}

The first approach solves the system in~\eqref{eq:system-moments-alpha} directly, ignoring the variance of the second variable given that, with probability 1, no solution would exist otherwise.
Therefore, the method of moments 1 (MM1) estimator $\hat{\boldsymbol{\alpha}} = (\hat\alpha_1, \hat\alpha_2, \hat\alpha_3, \hat\alpha_4)$ for $\boldsymbol{\alpha}$ is
\begin{gather}
    \label{eq:MM1}
    \begin{aligned}
    \tilde{\alpha}_4 &= \bar{\alpha}\left(P_n\sqrt{\bar{X}_n\bar{Y}_n(1-\bar{X}_n)(1-\bar{Y}_n)} + (1-\bar{X}_n)(1-\bar{Y}_n)\right) \\
    \hat{\alpha}_1 &= \max\{0, (\bar{X}_n + \bar{Y}_n - 1)\bar{\alpha} + \tilde{\alpha}_4\}, \\
    \hat{\alpha}_2 &= \max\{0,  (1 - \bar{Y}_n)\bar{\alpha} - \tilde{\alpha}_4\}, \\
    \hat{\alpha}_3 &= \max\{0, (1-\bar{X}_n)\bar{\alpha} - \tilde{\alpha}_4\}, \\
    \hat{\alpha}_4 &= \max\{0, \tilde{\alpha}_4 \},
    \end{aligned}
\end{gather}
where
\begin{equation*}
    \bar{\alpha} = \frac{\bar{X}_n - \bar{X}_n^2 - S_{X,n}^2}{S_{X,n}^2}.
\end{equation*}

Since the solution may be non-positive depending on the values of $\bar{X}_n, \bar{Y}_n, S_{X,n}^2$ and $P_n$, we set $\hat{\alpha}_i = 0$ as an approximation when this happens. 
~\cite{arnold2011flexible} use the same heuristic method for these situations. 

\begin{theoremEnd}{proposition}[Solution sign]
    Let $(\hat{\alpha}_1, \hat{\alpha}_2, \hat{\alpha}_3, \hat{\alpha}_4)$ be the solution given by~\eqref{eq:system-solution} with the observed moments $\hat{m}_1, \hat{m}_2, \hat{v}_1$ and $\hat\rho$. 
    If $\hat{v}_1 < \hat{m}_1(1-\hat{m}_1)$, then at most one coordinate is non-positive.
\end{theoremEnd}

\begin{proofEnd}
    Since $\bar{\alpha} > 0$ by hypothesis, define $\hat{b}_i = \hat{\alpha}_i/\bar{\alpha}$ and then notice the following relations, which are valid by consequence of the definition of the method of moments solution:
    \begin{enumerate}
        \item $\hat{\beta}_1 + \hat{\beta}_2 = \dfrac{\hat{\alpha}_1 + \hat{\alpha}_2}{\bar{\alpha}} = \bar{X}_n > 0$;
        \item $\hat{\beta}_1 + \hat{\beta}_3 = \dfrac{\hat{\alpha}_1 + \hat{\alpha}_3}{\bar{\alpha}} = \bar{Y}_n > 0$;
        \item $\hat{\beta}_2 + \hat{\beta}_4 = 1 - \dfrac{\hat{\alpha}_1 + \hat{\alpha}_3}{\bar{\alpha}} = 1 - \bar{Y}_n > 0$;
        \item $\hat{\beta}_3 + \hat{\beta}_4 = 1 - \dfrac{\hat{\alpha}_2 + \hat{\alpha}_3}{\bar{\alpha}} = 1 - \bar{X}_n > 0$;
        \item $\hat{\beta}_1 + \hat{\beta}_4 = \dfrac{\hat{\alpha}_1 + \hat{\alpha}_4}{\bar{\alpha}} = 2P_n\sqrt{\bar{X}_n\bar{Y}_n(1-\bar{X}_n)(1-\bar{Y}_n)} + \bar{X}_n\hat{m_2} + (1-\bar{X}_n)(1-\bar{Y}_n) > 0$, by considering that $P_n > -1$ and minimising the remaining expression. 
        Moreover, by considering that $P_n < 1$, we achieve that $\hat{\beta}_1 + \hat{\beta}_4 < 1$, which implies that $\hat{\beta}_2 + \hat{\beta}_3 > 0$.
    \end{enumerate}

    We can thus conclude that if $\hat{\beta}_j \le 0$ for some $j$, then $\hat{\beta}_i > 0$ for $i \neq j$.
    In other words, at most one coordinate is non-positive.
\end{proofEnd}
The proposition above implies that MM1 will have at most one zero value.
After estimating $\hat\parameter$ from $\dd$ using MM1, \underline{if it is strictly positive}, we can set the random pair $(\tilde{X}, \tilde{Y}) \sim \operatorname{BivariateBeta}(\hat\parameter)$.
Then,
\[
\var(\tilde{X}) = \frac{\ev[\tilde{X}](1-\ev[\tilde{X}])}{s_{\alpha} + 1} \implies s_{\alpha} + 1 = \frac{\ev[\tilde{X}](1-\ev[\tilde{X}])}{\var(\tilde{X})}.
\]
Therefore, 
\[
\var(\tilde{Y}) = \var(\tilde{X})\frac{\ev[\tilde{Y}](1-\ev[\tilde{Y}])}{\ev[\tilde{X}](1-\ev[\tilde{X}])},
\]
which gives a formula for the variance of the second variable since the estimate does not use its sample counterpart.

\subsubsection*{Method of moments with the analytical solution for three equations}

The second estimator we consider is based on~\autoref{prop:system-three-moments}.
For a given value of $\alpha_4$, observe that the expression in~\eqref{eq:system-three-solution} gives
\[
s_{\alpha} = \sum_{j=1}^4 \alpha_j = \frac{\alpha_4}{(1 - \bar{X}_n)(1 - \bar{Y}_n) + P_n\sqrt{\bar{X}_n \bar{Y}_n (1-\bar{X}_n)(1-\bar{Y}_n)}}.
\]
Moreover, we would like that 
\begin{equation}
    \label{eq:relation-v1v2}
    S_{X,n}^2 = \frac{\bar{X}_n(1 - \bar{X}_n)}{s_{\alpha}+1}, S_{Y,n}^2 = \frac{\bar{Y}_n(1 - \bar{Y}_n)}{s_{\alpha}+1} \implies s_{\alpha} + 1 = \frac{\bar{X}_n(1 - \bar{X}_n)}{S_{X,n}^2} = \frac{\bar{Y}_n(1 - \bar{Y}_n)}{S_{Y,n}^2}.
\end{equation}
Therefore, we minimise the expression 
\[
{\left(s_{\alpha} + 1 - \frac{\bar{X}_n(1 - \bar{X}_n)}{S_{X,n}^2}\right)}^2 + {\left(s_{\alpha} + 1 - \frac{\bar{Y}_n(1 - \bar{Y}_n)}{S_{Y,n}^2}\right)}^2, 
\]
for which the solution is 
\[
\tilde{\alpha_4} = \left((1 - \bar{X}_n)(1 - \bar{Y}_n) + P_n\sqrt{\bar{X}_n \bar{Y}_n (1-\bar{X}_n)(1-\bar{Y}_n)}\right)\left(\frac{\frac{\bar{X}_n(1 - \bar{X}_n)}{S_{X,n}^2} + \frac{\bar{Y}_n(1 - \bar{Y}_n)}{S_{Y,n}^2}}{2} - 1\right).
\]
Then, the method of moments 2 (MM2) estimator $\hat{\boldsymbol{\alpha}}$ for $\boldsymbol{\alpha}$ is
\begin{equation}
  \label{eq:mm2}
  \begin{aligned}
    \hat\alpha_1 &= \max\left\{0,  \tilde\alpha_4\frac{\bar{X}_n\bar{Y}_n + P_n\sqrt{\bar{X}_n\bar{Y}_n(1-\bar{X}_n)(1-\bar{Y}_n)}}{(1-\bar{X}_n)(1-\bar{Y}_n) + P_n\sqrt{\bar{X}_n\bar{Y}_n(1-\bar{X}_n)(1-\bar{Y}_n)}}\right\}, \\
    \hat\alpha_2 &= \max\left\{0, \tilde\alpha_4\frac{\bar{X}_n(1-\bar{Y}_n) - P_n\sqrt{\bar{X}_n\bar{Y}_n(1-\bar{X}_n)(1-\bar{Y}_n)}}{(1-\bar{X}_n)(1-\bar{Y}_n) + P_n\sqrt{\bar{X}_n\bar{Y}_n(1-\bar{X}_n)(1-\bar{Y}_n)}}\right\}, \\
    \hat\alpha_3 &= \max\left\{0, \tilde\alpha_4\frac{\bar{Y}_n(1-\bar{X}_n) - P_n\sqrt{\bar{X}_n\bar{Y}_n(1-\bar{X}_n)(1-\bar{Y}_n)}}{(1-\bar{X}_n)(1-\bar{Y}_n) + P_n\sqrt{\bar{X}_n\bar{Y}_n(1-\bar{X}_n)(1-\bar{Y}_n)}}\right\}, \\
    \hat\alpha_4 &= \max\{0, \tilde\alpha_4\}.
  \end{aligned}
\end{equation}
    
\begin{remark}
    Notice that we have a necessary and sufficient condition on $\bar{X}_n, \bar{Y}_n$ and $P_n$ for the estimator  to be strictly positive, as presented in~\autoref{sec:moments}.
    Moreover, let  
    \[
    D = (1-\bar{X}_n)(1-\bar{Y}_n) + P_n\sqrt{\bar{X}_n\bar{Y}_n(1-\bar{X}_n)(1-\bar{Y}_n)}.
    \]
    Considering~\autoref{remark:inequality-beta}, it is clear that $\tilde{\alpha} > 0 \iff D > 0$. 
    If $P_n > 0$, we have $D > 0$. 
    Then $\tilde{\alpha}_4 > 0$ and, consequently, $\alpha_1 > 0$ and $\alpha_4 > 0$. 
    Moreover, if $P_n < 0$, then $\hat{\alpha}_2 > 0$ and $\hat{\alpha}_3 >0$, because $\tilde{\alpha}_4/D > 0$. 
    Therefore, the distribution $\operatorname{BivariateBeta}(\hat{\boldsymbol{\alpha}})$ might not be defined since $\hat{\alpha}_i$ may be zero but the moments will still be well-defined.
\end{remark}    

\subsubsection*{Method of moments with the analytical solution for two equations}

The third possibility for the moment estimator worth exploring is to use relations~\eqref{eq:alpha1-as-function-alpha3-alpha4} and~\eqref{eq:alpha2-as-function-alpha3-alpha4} to define $\hat{\alpha}_1$ and $\hat{\alpha}_2$ as functions of $\hat{\alpha}_3$ and $\hat{\alpha}_4$ from the values of $\bar{X}_n$ and $\bar{Y}_n$.
To find values for $\hat{\alpha}_3$ and $\hat{\alpha}_4$, we use the expressions in~\eqref{eq:relation-v1v2} and 
\[
\rho = \frac{\alpha_1\alpha_4 - \alpha_2\alpha_3}{\sqrt{(\alpha_1+\alpha_2)(\alpha_1+\alpha_3)(\alpha_2+\alpha_4)(\alpha_3+\alpha_4)}} = \frac{(1-m_1)((\alpha_3+\alpha_4)m_2 - \alpha_3)}{(\alpha_3+\alpha_4)\sqrt{m_1m_2(1-m_1)(1-m_2)}},
\]
which gives the expression
\[
\frac{\alpha_3}{\alpha_3 + \alpha_4} = m_2 -  \frac{\rho\sqrt{m_1m_2(1-m_1)(1-m_2)}}{1-m_1} := E \implies (E-1)\alpha_3 + E\alpha_4 = 0.
\]

Therefore, we define $\hat{\alpha}_3$ and $\hat{\alpha}_4$ to be the solution to the problem 
\begin{equation}
    \label{eq:minimisation-mm3}
    \begin{split}
            \min_{\alpha_3, \alpha_4 > 0} &{\left(s_{\alpha} - \dfrac{\bar{X}_n(1 - \bar{X}_n)}{S_{X,n}^2} - 1\right)}^2 + {\left(s_{\alpha} - \dfrac{\bar{Y}_n(1 - \bar{Y}_n)}{S_{Y,n}^2}-1\right)}^2 + {((E-1)\alpha_3 + E\alpha_4)}^2, \\
            \text{s.t. } &(\bar{X}_n + \bar{Y}_n - 1)\alpha_3 + \bar{Y}_n \alpha_4 > 0, \\
            &(1 - \bar{Y}_n)\alpha_3 + (\bar{X}_n - \bar{Y}_n)\alpha_4 > 0,
    \end{split}
\end{equation}
where $s_{\alpha} = (\alpha_3 + \alpha_4)/(1 - \bar{X}_n)$.
This defines the method of moments 3 (MM3) estimator for $\parameter$.

\subsubsection*{Method of moments with no analytical solution}

Finally, we consider the approach defined by~\cite{olkin2015constructions}, which seeks to minimise the following expression: 
\begin{equation}
    \label{eq:olkin-method}
    \begin{split}
    {\left(\bar{X}_n - \frac{\alpha_1 + \alpha_2}{s_{\alpha}}\right)}^2 + {\left(\bar{Y}_n - \frac{\alpha_1 + \alpha_3}{s_{\alpha}}\right)}^2 &+ {\left(P_n - \frac{\alpha_1 \alpha_4 - \alpha_2 \alpha_3}{\sqrt{(\alpha_1 + \alpha_2)(\alpha_1 + \alpha_3)(\alpha_2 + \alpha_4)(\alpha_3 + \alpha_4)}}\right)}^2 +\\
    {\left(S_{X,n}^2 - \dfrac{(\alpha_1+\alpha_2)(\alpha_3+\alpha_4)}{s_{\alpha}^2(s_{\alpha}+1)}\right)}^2 &+ {\left(S_{Y,n}^2 - \dfrac{(\alpha_1+\alpha_3)(\alpha_2+\alpha_4)}{s_{\alpha}^2(s_{\alpha}+1)}\right)}^2,
    \end{split}
\end{equation}
subject to $\alpha_1, \alpha_2, \alpha_3, \alpha_4 > 0$. 
Moreover,~\cite{olkin2015constructions} consider the additional restriction 
\begin{equation}
    \label{eq:olkin-inequality}
    s_{\alpha} \le \max\left(\frac{\bar{X}_n(1-\bar{X}_n)}{S_{X,n}^2},  \frac{\bar{Y}_n(1-\bar{Y}_n)}{S_{Y,n}^2}\right)-1,
\end{equation}
based on the expression in~\eqref{eq:relation-v1v2}. 
The solution $\hat{\boldsymbol{\alpha}}$ to the minimisation problem~\eqref{eq:olkin-method} is the method of moments 4 (MM4) estimator.

\begin{remark}
    We implicitly defined the quadratic loss for each minimisation strategy given it is differentiable and computationally inexpensive.
    Nevertheless, other losses can be used, such as the absolute, the absolute percentage or the weighted quadratic, depending on the application.
\end{remark}

In short, we have described four approaches: 

\begin{enumerate}
    \item[(MM1)] Solve four equations according to~\autoref{prop:solution-to-system-bivariate-beta} and if the solution is non-positive, return $0$ as an heuristic approach~\citep{arnold2011flexible} --- at most one coordinate.
    \item[(MM2)] Solve three equations involving ($m_1, m_2, \rho$) and choose $\alpha_4$ to minimise the relations compared to $v_1$ and $v_2$.
    If the solution is negative, use the same approach as MM1. 
    \item[(MM3)] Solve two equations involving ($m_1, m_2$) and the optimisation problem in~\eqref{eq:minimisation-mm3}
    \item[(MM4)] Minimise the expression in~\eqref{eq:olkin-method} subject to $\alpha_i > 0$ and inequality~\eqref{eq:olkin-inequality}.
\end{enumerate}

\subsection{Bayesian estimation}\label{sec:bayesian_estimation}

By defining a prior distribution for $\parameter$ and using the density in~\eqref{eq:dist-X-Y}, Bayes rule updates our knowledge about the parameter conditional on $\dd$ through the posterior distribution, $p(\parameter | \dd)$.
After setting a loss function, we can derive the Bayes estimator $\hat{\boldsymbol\alpha}$ by minimising the expected posterior loss.
Here we use the posterior mean (quadratic loss) and median (absolute loss) as point estimation strategies, which we shall call BE1 and BE2, respectively.
The density $p(\parameter | \dd)$ is not tractable, and therefore one needs to resort on numerical methods to compute EB1 and EB2.
We employ dynamic Hamiltonian Monte Carlo (HMC) as implemented in the Stan programming language~\cite[]{stan} to obtain approximate posterior samples.

Since the joint density of the data (i.e.\ the likelihood,~\autoref{eq:dist-X-Y}) depends on a tricky integration step (see~\autoref{sec:integration_comment}), direct application of the HMC algorithm is not feasible.
We circumvent this by including a latent variable $u \in \R^n$ in the model.
Considering the density of the Dirichlet distribution given in~\eqref{eq:dirichlet-distribution} and setting $(u,x,y) = (u_1, u_1 + u_2, u_1 + u_3)$, by the Change of Variables formula, 
\[
f(u, x, y | \boldsymbol\alpha) = \frac{1}{B(\boldsymbol{\alpha})}u^{\alpha_1-1}{(x-u)}^{\alpha_2-1}{(y-u)}^{\alpha_3-1}{(1-x-y+u)}^{\alpha_4-1}\one\{u \in \Omega\}.
\]
Writing the \textit{complete likelihood} function $L(\boldsymbol{\alpha}, u| x,y) = f(x,y|\boldsymbol{\alpha}, u)$ and using the conditional density expression, we get the posterior 
\begin{equation*}
    \begin{split}
        p(\boldsymbol\alpha, u | x,y) &\propto \prod_{i=1}^n L(\boldsymbol{\alpha}, u_i| x_i, y_i)f(u_i|\boldsymbol{\alpha})\pi(\boldsymbol\alpha), \\
        p(\boldsymbol\alpha, u | x,y) &\propto \prod_{i=1}^n \frac{f(u_i, x_i, y_i | \boldsymbol\alpha)}{f(u_i|\boldsymbol{\alpha})}f(u_i|\boldsymbol{\alpha})\pi(\boldsymbol\alpha), \\
        &= \prod_{i=1}^n f(u_i,x_i,y_i|\boldsymbol{\alpha})\pi(\boldsymbol\alpha).
    \end{split}
\end{equation*}

\begin{figure}
    \centering
    \includegraphics[width=0.7\textwidth]{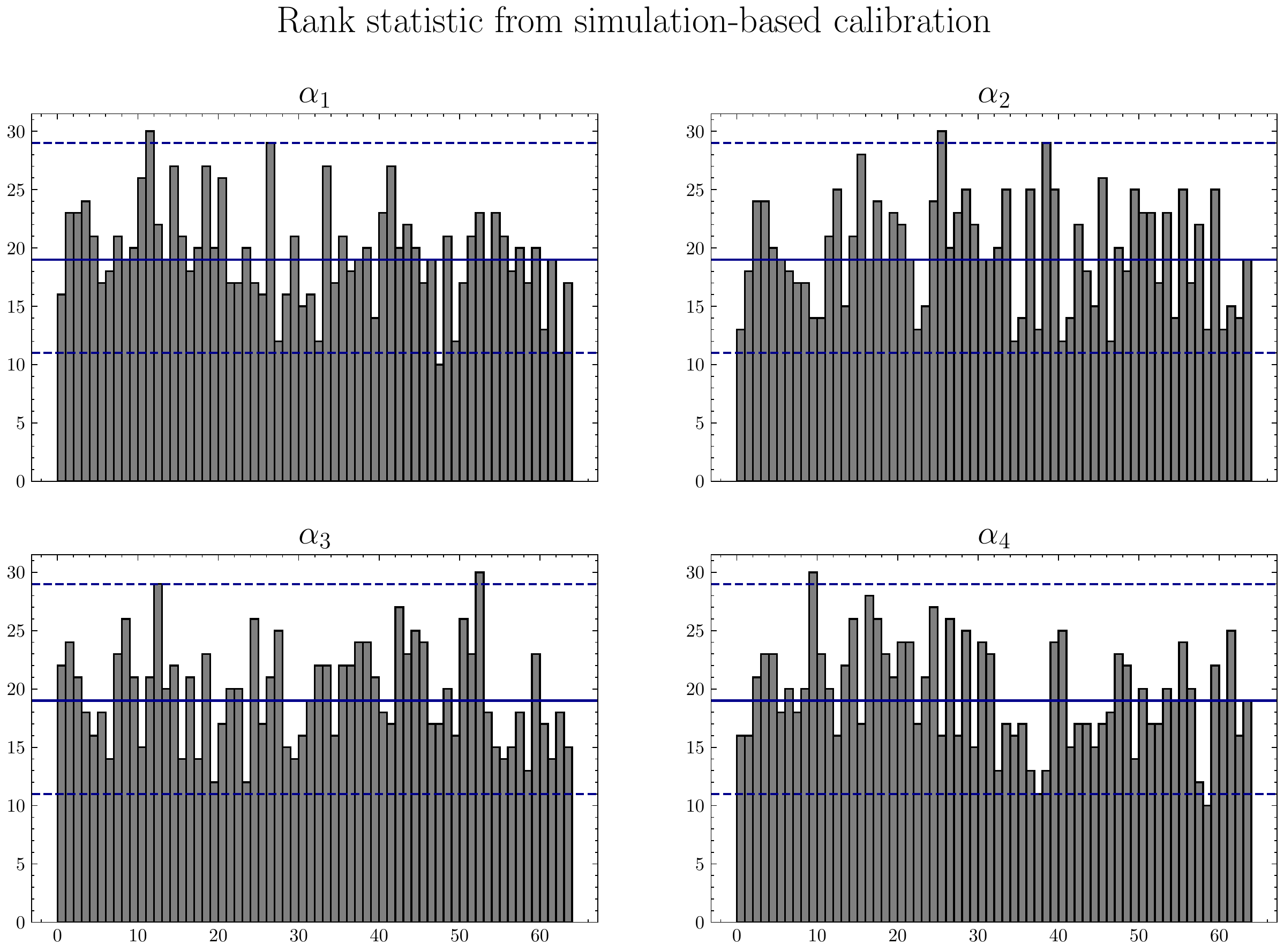}
    \caption{{\bf Simulation-based calibration:} Histogram of the rank statistic $r$ in~\eqref{eq:rank_stat} for each component of $\parameter$.
    The region between the dashed blue lines indicates the 95\% interval under a discrete uniform distribution and the solid blue line is the median.
    We set the number of bins to $L=63$ and perform $N=1252$ experiments.}\label{fig:sbc_rank_statistic}
\end{figure}

In order to validate the computed posterior distribution, we use a {\em simulation-based calibration\/} (SBC) approach, which is based on the histogram of the rank statistics~\cite[]{talts2018validating}.
In this method, we use the fact that if $\tilde{\parameter} \sim \pi(\parameter)$, $\tilde{\dd} \sim \operatorname{BivariateBeta}(\tilde{\parameter})$ and $\{\parameter^1, \dots, \parameter^L\} \sim p(\parameter|\tilde{\dd})$, the rank statistic
\begin{equation}
    \label{eq:rank_stat}
    r(\{\alpha^1_i, \dots, \alpha^L_i\}, \tilde{\alpha}_i) = \sum_{l=1}^L 1[\alpha^l_i < \tilde{\alpha}_i]
\end{equation}
is uniformly distributed over the integers in $[0, L]$ for each $i=1,\dots,4$.
Using independent $\operatorname{Gamma}(1,1)$ for the prior distributions leads to the result in~\autoref{fig:sbc_rank_statistic}.
One drawback of this formulation is that the symplectic integrator inside the HMC algorithm (leapfrog) presents divergences when $\parameter$ is small, as shown in~\autoref{fig:sbc_divergences}.
Besides SBC, other MCMC and HMC-specific diagnostics are relevant, such as the $\hat{R}$~\citep{gelman1992inference}, the presence of divergences, the energy~\cite[]{betancourt2016diagnosing} and the effective sample size (ESS).
All diagnostics are available in the notebooks provided in~\url{https://github.com/lucasmoschen/bivariate-beta}.
Furthermore, for the numerical experiments, we used $2,000$ warmup iterations, $2,000$ sampling iterations and an {\tt adapt\_delta} of $0.9$.
The other computational specifications are the default in Stan.

\begin{remark}
    Notice that $(x_i, y_i)$ given $u_i$ and $\boldsymbol\alpha$ are independent of other values $x_j, y_j, u_j$ for $j \neq i$, which allows for efficient vectorised implementations.  
    Despite having increased the parameter dimension from $4$ to $n+4$, the execution time was reduced by a factor of $100$ when compared to integrating via quadrature in Stan at each iteration.
\end{remark}

\subsubsection{Prior distributions for \texorpdfstring{$\parameter$}{}}

A natural question to ask is which prior distribution should one place on $\parameter$.
We consider two families of proper distributions to specify $\pi(\parameter)$ considering \textit{a priori} independent coordinates.

\begin{enumerate}[(a)]
    \item $\{\alpha_i\}$ are gamma-distributed with scale parameter $a_i$ and rate parameter $b_i$. 
    For simplicity, here we set $a_i = a$ and $b_i = b$ in this paper.
    This should be the case unless different information is provided for each $\alpha_i$;
    
    \item $\{\alpha_i\}$ have a distribution with density
    \[
    \pi(\alpha_i) = \frac{p}{C}\one\{\alpha_i \le C\} + (1-p)\lambda e^{-\lambda(\alpha_i-C)}\one\{\alpha_i > C\}.
    \]
    This works as a proper approximation to the uniform distribution on $[0, +\infty)$, which is improper.
    This prior thus encodes a situation where we have a uniform distribution when $\alpha_i \in [0, C]$ with probability $p$ and an exponential with rate parameter $\lambda$ when $\alpha_i > C$ with probability $1-p$. 
    The value of $\lambda$ is chosen to ensure that the density is continuous at $\alpha_i = C$.
\end{enumerate}

\begin{figure}[!htbp]
    \centering
    \includegraphics[width=0.6\textwidth]{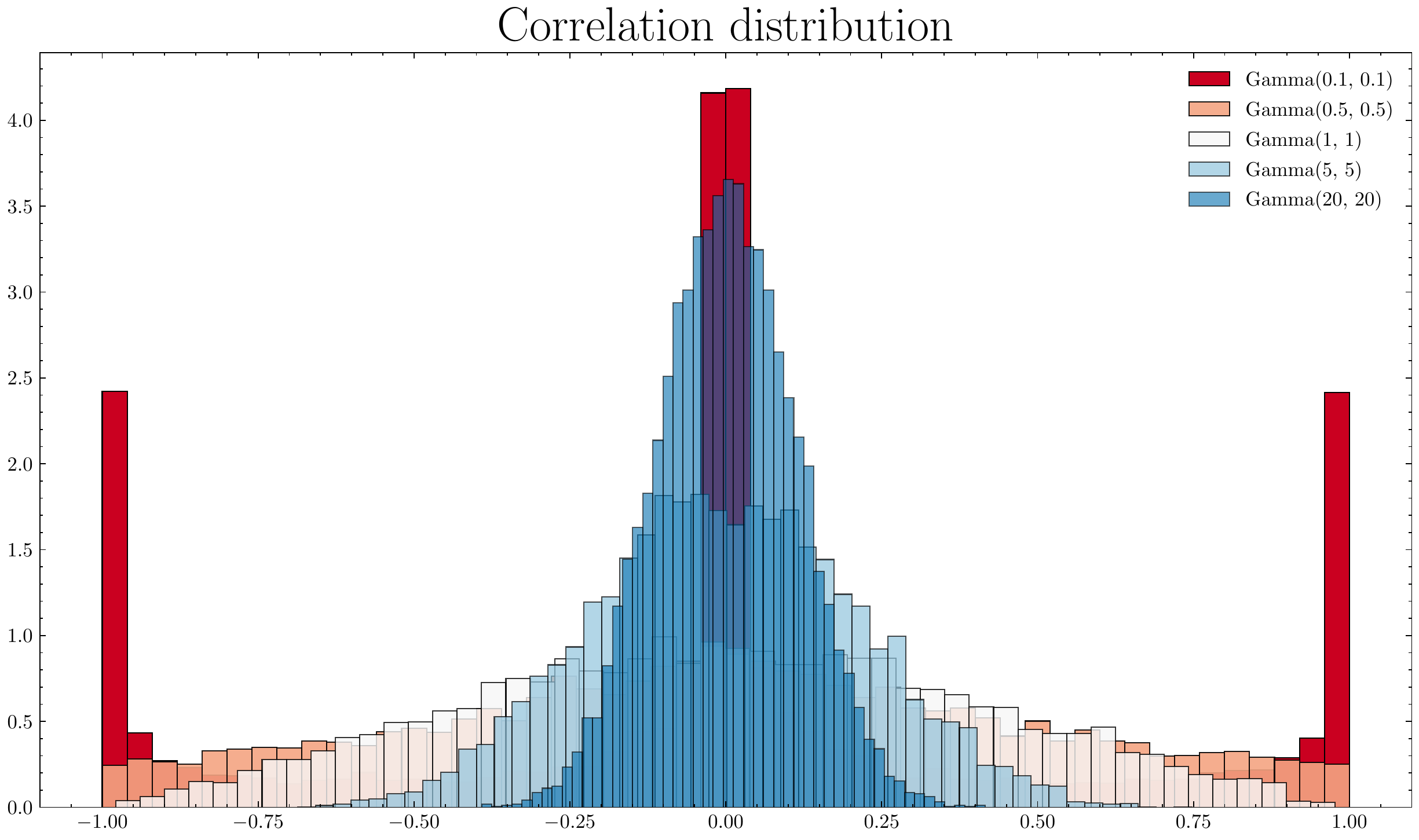}
    \caption{{\bf Distribution of the correlation induced by different priors:} Distribution of $\rho = \cor(X,Y)$ when $\alpha_1, \dots, \alpha_4 \overset{iid}{\sim} \operatorname{Gamma}(a,a)$ for different values of $a$.}\label{fig:correlation-distribution-gamma-a-a}
\end{figure}

We employ prior predictive checking (see e.g.~\cite{gabry2019visualization}) to understand the implications of each prior choice. 
Some interesting observations are:
\begin{enumerate}[(a)]
    \item Using the same distribution for each $\alpha_i$ generates a symmetric distribution for the correlation $\rho$, as~\autoref{fig:correlation-distribution-gamma-a-a} shows.
    In order to incorporate \textit{a priori} information about the sign of $\rho$, one needs to have a gamma distribution with a larger or lower mean for $\alpha_1$ and $\alpha_4$ when compared to $\alpha_2$ and $\alpha_3$.
    Moreover, in order to have different distributions for the means of $X$ and $Y$, the distributions of $\alpha_2$ and $\alpha_3$ have to be distinct.
    
    \item When the uniform-exponential distribution is set as prior, small and high values for $C$ are not good.
    The former case leads to bimodal distributions, while the latter induces a distribution for the variances of $X$ and $Y$ that is too concentrated.
    Furthermore, changes in the values of $C$ and $p$ appear to have little impact on the correlation induced prior.
\end{enumerate}
For more details, please see the Supplementary Material. 

\section{Elicitation of a bivariate beta prior}\label{sec:elicitation}

An important application of the bivariate beta construction is to act as the prior distribution for variables on ${[0, 1]}^2$ that are believed or known to be correlated \textit{a priori}, such as the sensitivity and specificity of a diagnostic test.
In this section we briefly discuss how one can elicit a bivariate beta, i.e., specify the value of $\parameter$ by reasoning about its properties such as moments and tail probabilities.

Let $X_1, \dots, X_m$ be a random sample with distribution $P_{\theta}$ and suppose that the domain of the model parameter $\theta = (\theta_1, \theta_2)$ is ${[0,1]}^2$.
Here, we suppose that $\theta \sim \operatorname{BivariateBeta}(\parameter)$ and propose an elicitation approach for $\parameter$.
To identify the parameter $\parameter$, we consider that the researcher wants to encode the values $m_1 = \ev[\theta_1], m_2 = \ev[\theta_2], v_1 = \var(\theta_1), v_2 = \var(\theta_2)$ and $\rho = \cor(\theta_1, \theta_2)$ in the distribution of $\parameter$.
By inequality~\eqref{eq:inequality-bivariate-beta} from~\autoref{prop:solution-to-system-bivariate-beta}, we already know if we can define a bivariate beta distribution with these moments. 
Therefore, we set the following strategy:

\begin{enumerate}[(a)]
    \item If the solution in~\eqref{eq:system-solution} belongs to the parameter space, we use it as the parameter for the prior. 
    
    \item Otherwise, if we give less importance to $v_1$ than $m_1, m_2$ and $\rho$, we use the MM2 estimate if this approach leads to a solution in the parameter space;
    
    \item Finally, if the above methods are not convenient, MM3 and MM4 should be used since they have a solution in the parameter space independent of the input values. If the information about $m_1$ and $m_2$ is more robust, MM3 is preferable, otherwise MM4 provides a compelling option.
\end{enumerate}

Although we defined information about marginal means and variances as inputs, each pair of information about the marginals can used transforming into the mean and the variance. 
For instance, we can define the mean and a quantile and convert this information to mean and variance of a marginal variable.

\begin{remark}
    As noticed in~\autoref{sec:theory}, the sum of the marginal parameters equals $s_{\alpha}$ for both $X$ and $Y$.
    An interpretation of the beta parameters accounts for the number of successes and failures before the actual experiment. Therefore, the sum of the parameters is the number of trials or {\em pseudo-trials}.
    In that sense, the number of pseudo-trials we use is the same for both $X$ and $Y$.
\end{remark}

\section{Numerical experiments}

We use $1,000$ Monte Carlo simulations to compare the six estimation strategies for $\parameter$ (four method-of-moments and two Bayes estimates) by measuring bias, mean squared error (MSE), mean absolute percentage error (MAPE), runtime and coverage for the interval estimates.
All experiments were run on a Linux PC with an Intel (R) Core (TM) i7--1165G7 2.8GHz processor (4 cores) and 16 GB of memory.
All computer code for reproducing the experiments and using the proposed methods can be found under a permissive licence at~\url{https://github.com/lucasmoschen/bivariate-beta}.

\subsection{Recovering parameters under the bivariate beta}\label{sec:recovering-bivariate-beta}

We begin our investigation under the \underline{well-specified} case, where the data come from the bivariate beta model under consideration.
Let $(X,Y) \sim \operatorname{BivariateBeta}(\boldsymbol\alpha)$, where $\boldsymbol\alpha$ is an unknown parameter.
For this experiment, we set three representative sets of parameters: $\boldsymbol\alpha = (1,1,1,1)$, $\boldsymbol\alpha = (2, 7, 3, 1)$ and $\boldsymbol\alpha = (0.7, 0.9, 2.0, 1.5)$. 
We then generate $1,000$ simulations for each sample size of $n=50$ and $n=200$ samples. 
One can then compute estimates for bias, MSE and MAPE through Monte Carlo.
For the method of moments, we perform a non-parametric bootstrap estimate to get the 95\% confidence interval with $B=500$ bootstrap samples and thereafter estimate coverage.
The numerical results are summarised in Tables~\ref{tab:alpha-experiment1},~\ref{tab:alpha-experiment2} and~\ref{tab:alpha-experiment3} in~\autoref{appendix:tables}.

\begin{figure}[!htpb]
    \centering
    \includegraphics[width=0.6\textwidth]{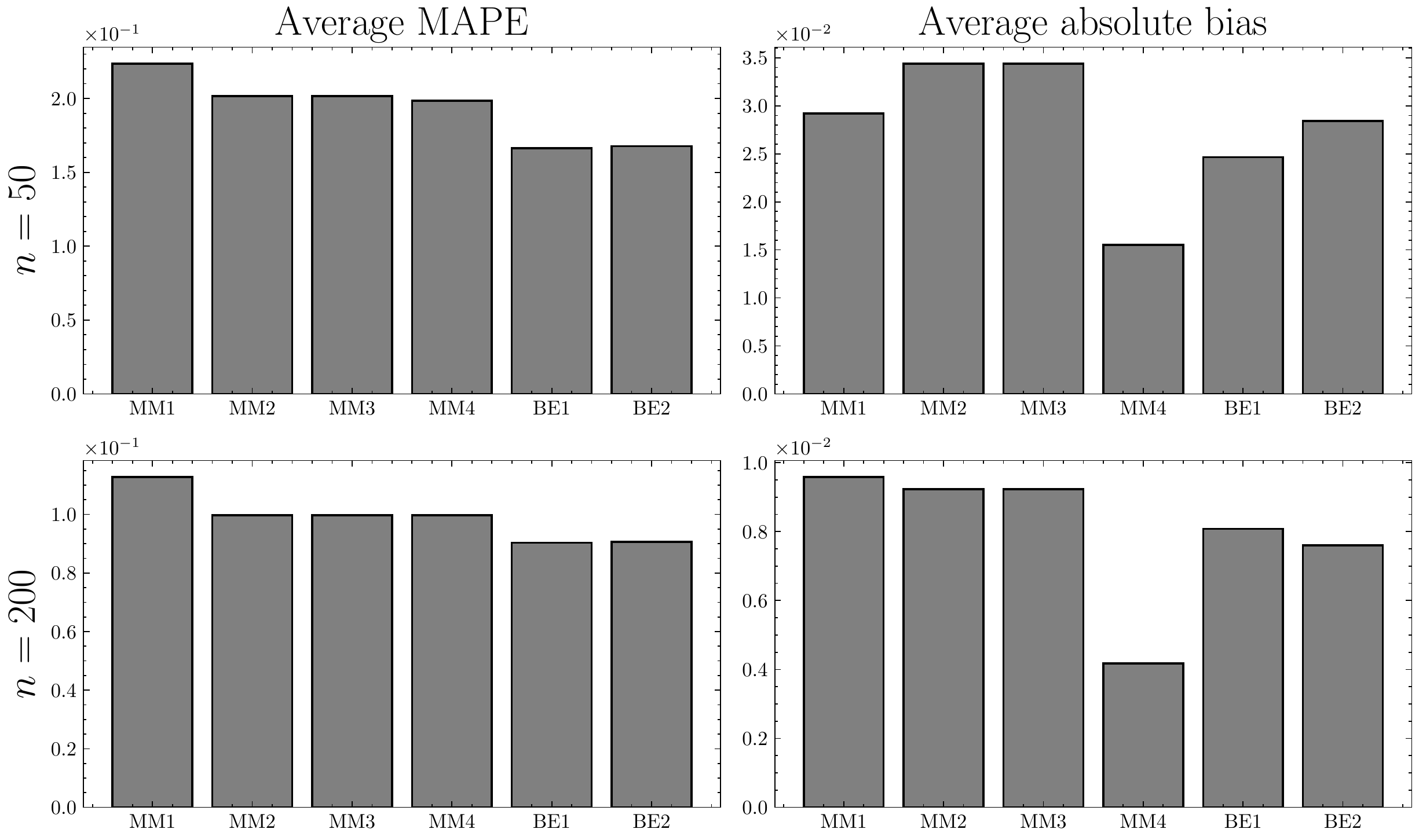}
    \caption{{\bf Estimation performance} when the true value is $\boldsymbol{\alpha} = (0.7,0.9,2,1.5)$.
    The average MAPE is the mean of the MAPEs for each estimate $\hat\alpha_i$, while the average absolute bias considers the mean of the absolute values of the bias of each estimate.
    }\label{fig:comparing_methods_mape_bias2}
\end{figure}

\begin{figure}[htbp]
    \centering
    \includegraphics[width=0.9\textwidth]{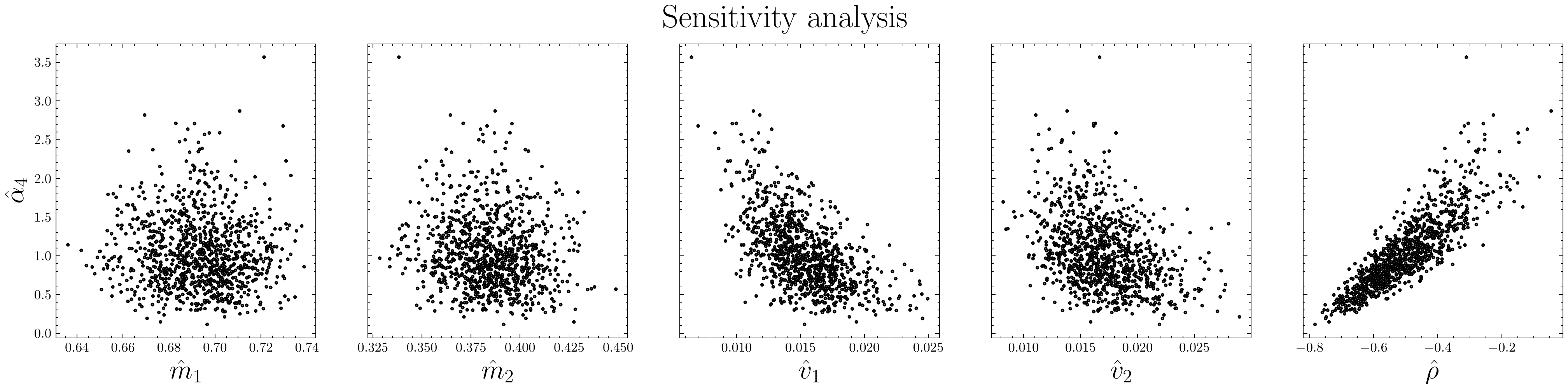}
    \caption{{\bf Estimates of $\hat{\alpha}_4$} for different sample datasets of size $n=50$ and $\alpha=[2,7,3,1]$ according to MM1. 
    We see the scatter plots of the sample moments of each dataset against the estimated value of $\hat{\alpha}_4$.
    We observe that $\hat{\rho}$ drives the most part of the variability of $\hat{\alpha}_4$.}\label{fig:alpha4_vs_moments}
\end{figure}

\begin{figure}[htbp]
    \centering
    \includegraphics[width=0.9\textwidth]{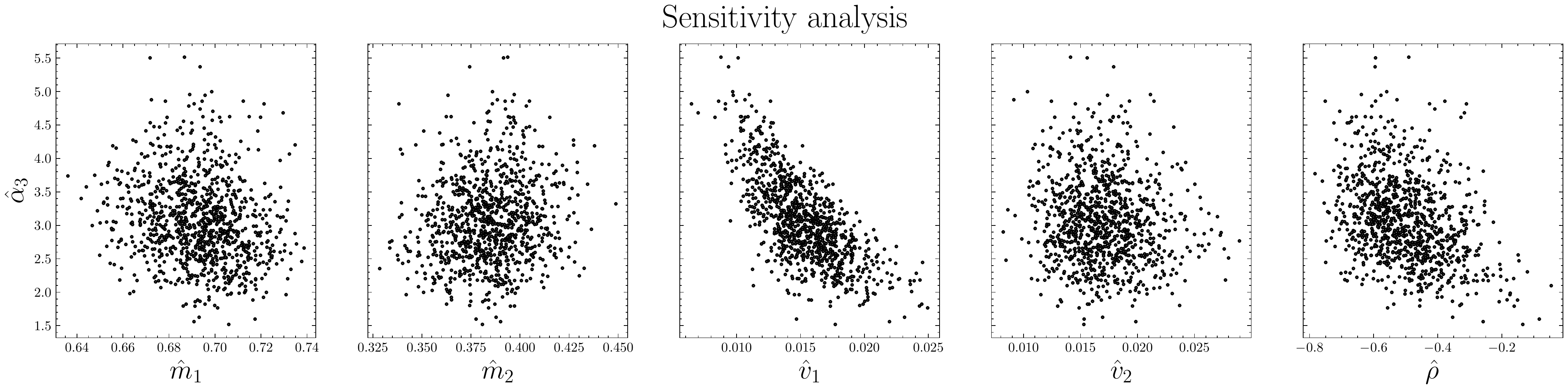}
    \caption{{\bf Estimates of $\hat{\alpha}_3$} for different sample datasets of size $n=50$ and $\alpha=[2,7,3,1]$ according to MM1. 
    We see the scatter plots of the sample moments of each dataset against the estimated value of $\hat{\alpha}_3$.
    We observe that $\hat{v}_1$ drives the most part of the variability of $\hat{\alpha}_3$.}\label{fig:alpha3_vs_moments}
\end{figure}

Figures~\ref{fig:comparing_methods_mape_bias} and~\ref{fig:comparing_methods_mape_bias2} show that MM2, MM3 and MM4 estimators have similar MAPE, but the latter has lower average bias.
We also observe that the mean error decreases when $n$ grows, which is a consequence of the consistency of the method of moments, by the Law of Large Numbers and the solution in equation~\eqref{eq:system-solution} being a continuous function of $m_1, m_2, v_1$ and $\rho$.
MM1 is worse than the others for two reasons: it ignores the information brought by the second variable's variance, and it solves the equations exactly, which can suffer from noise in the sample moments, especially when $n$ is not large enough, as depicted in~\autoref{fig:alpha4_vs_moments} and~\autoref{fig:alpha3_vs_moments}. 
These figures show the estimates of $\hat\alpha_4$ and $\hat\alpha_3$, respectively, against the empirical moments' estimates for a specified $\parameter$. 
In particular, $\hat{\rho}$ and $\hat{v}_1$ seemed to have the largest influence on the estimates.
Moreover, despite analysing the confidence intervals for each component independently, we can produce confidence sets for the whole parameter vector.
In~\autoref{fig:bootstrap_samples_2d} we plot the pairwise estimates for $\parameter$ provided by the bootstrap method.
In particular we note that the estimates of $\alpha_1$ and $\alpha_4$ are correlated and so are the estimates of $\alpha_2$ and $\alpha_3$.

For the Bayesian estimates, we used independent $\operatorname{Gamma}(1,1)$ distributions as priors for all three specifications.
Both posterior mean and median were better in a matter of MAPE, but have larger biases, especially the posterior mean, which is expected for the Bayes estimator.
\autoref{fig:comparing_methods_mape_bias003} shows how large can the Bayesian estimator's bias be when the prior is far from the true value. 
For instance, we are saying that $\pr(\alpha_2 > 7) < 0.001$ prior to observing the data, despite the true value being $\alpha_2 = 7$.
The observed coverage of the confidence/credibility intervals was between $93.6\%$ and $96.3\%$, which is close to the nominal $95\%$.
We thus conclude that the bootstrap method and the posterior distribution produced good interval estimates. 
It is important to notice that the Bayesian credible intervals (BCI) do not need to have good frequency characteristics, but in this case they do display good long-run operating properties.
However, when $\parameter = (2,7,3,1)$, the BCI underestimate the coverage. 

\begin{figure}
    \centering
    \includegraphics[width=0.3\textwidth]{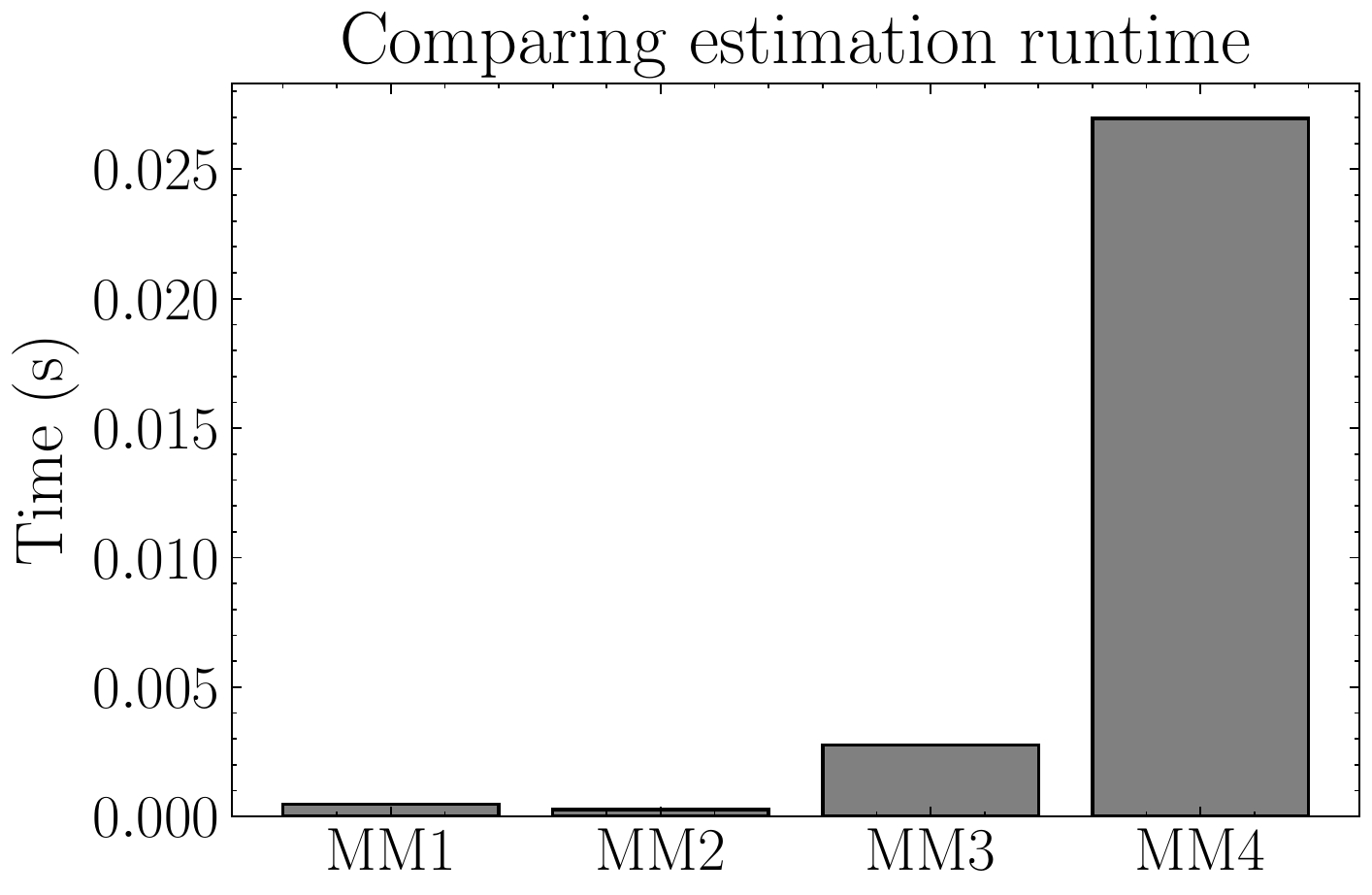}
    \caption{{\bf Method of moments runtime:} estimated runtime to execute the method of moments estimation in seconds.}\label{fig:runtime_moments_methods}
\end{figure}

Finally,~\autoref{fig:runtime_moments_methods} displays the difference between the estimated execution time for the method of moments.
Notice that MM1 and MM2 are more than 50 times faster than MM4, when $\parameter = (2,7,3,1)$ and $n=50$, given that they have an explicit formula, and MM4 depends on how close the initial value is from the minimum.
This pattern also occurs for the other specifications. 

\begin{figure}[!htbp]
    \centering
    \includegraphics[width=0.6\textwidth]{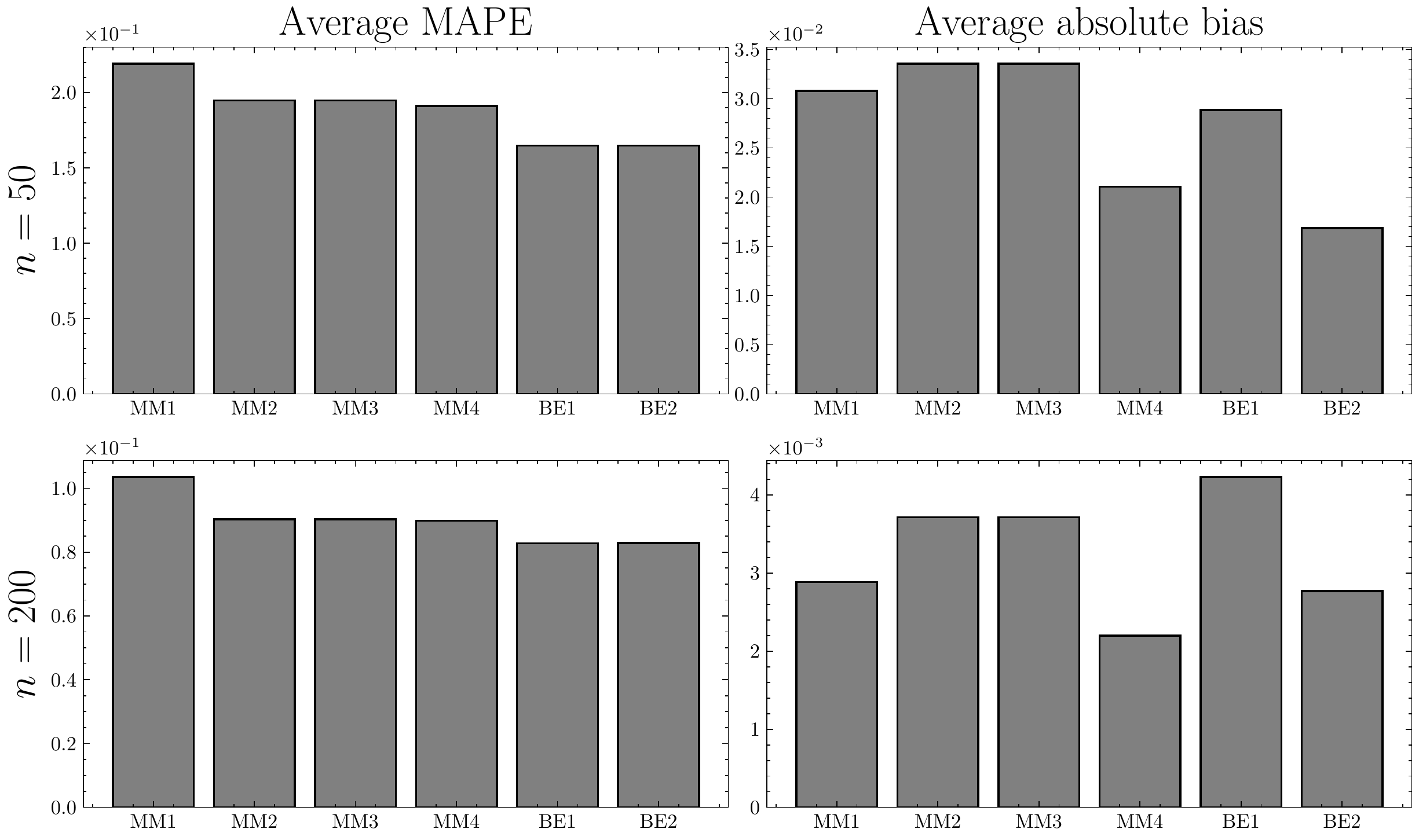}
    \caption{{\bf Estimation performance} when the true value is $\boldsymbol{\alpha} = (1,1,1,1)$.
    The average MAPE is the mean of the MAPEs for each estimate $\hat\alpha_i$, while the average absolute bias considers the mean of the absolute values of the bias of each estimate.
    }\label{fig:comparing_methods_mape_bias}
\end{figure}

\subsection{Model misspecification}\label{sec:misspecification}

We now move on to study how the methods proposed here perform when the bivariate model is misspecified, i.e.\ when the true data-generating process is some other distribution on ${[0, 1]}^2$.

Let $(X_1, Y_1), \dots, (X_n, Y_n)$ be i.i.d.\ random variables with a distribution whose unknown density $g(x,y)$ has support over ${[0,1]}^2$, and suppose we use the bivariate beta density $f_{\parameter}(x,y)$ to approximate $g$.
% The parameter estimation aims to find $\hat\parameter$ such that $f_{\hat\parameter}(\cdot)$ is close to $g(\cdot)$.
For the method of moments, we know that the solution in~\eqref{eq:system-solution} converges (almost surely) to $\hat\parameter$ such that $f_{\hat\parameter}$ has the same means, one marginal variance and correlation than $g$ if $\hat\parameter$ has positive coordinates.
Nonetheless, some issues might appear, such as (i) the empirical estimators of the moments given in equation~\eqref{eq:empirical-moments} may have high variances and thus yield bad estimates; (ii) the solution to the system may be negative, even for large values of $n$, given the results presented in~\autoref{fig:alpha-solutions}; and (iii) the moments of $g$, despite being well approximated, may not represent other characteristics of interest --- such as probabilities.

\begin{figure}[!htbp]
    \centering
    \includegraphics[width=0.3\textwidth]{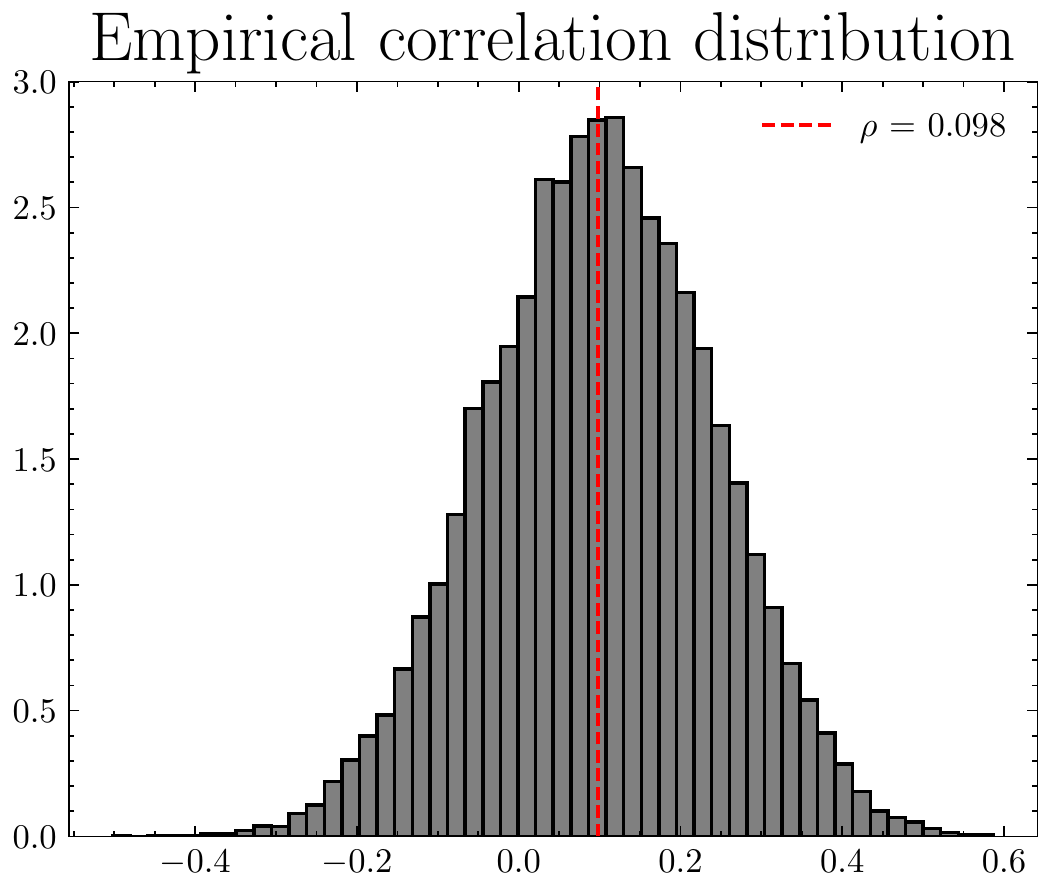}
    \caption{{\bf Sample distribution of the correlation}: sample distribution of the empirical correlation $P_n$, calculated through Monte Carlo samples when $\mu = (0,0)$, $\Sigma = [[1, 0.1], [0.1, 1]]$ and $n=50$.
    }\label{fig:sample_distribution_rho}
\end{figure}

\begin{figure}[!htbp]
    \centering
    \includegraphics[width=0.7\textwidth]{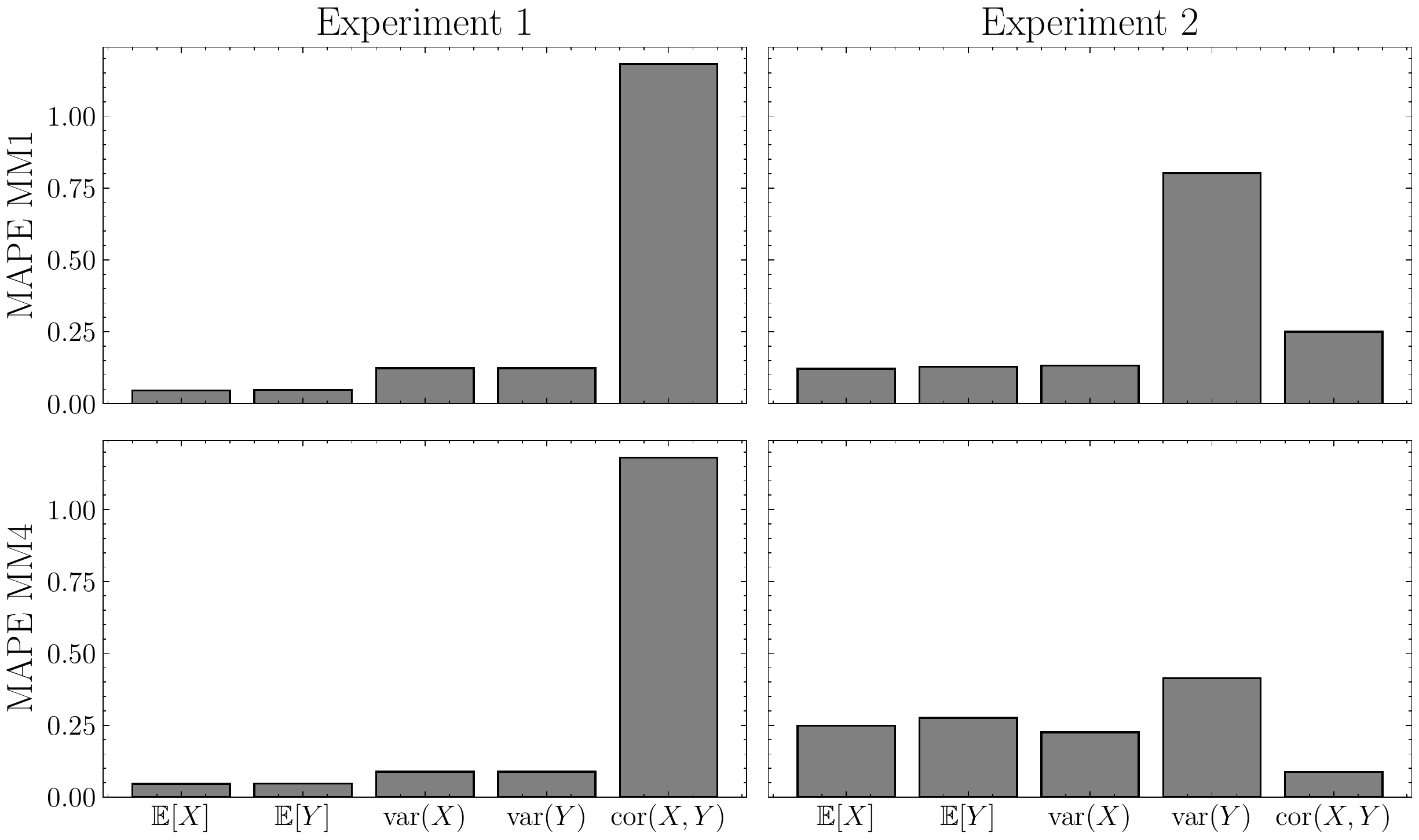}
    \caption{{\bf MAPE values} for the experiments in the misspecified case considering the true moments against the estimated through the bivariate beta with parameter $\hat\parameter$ obtained from MM1 and MM4.}\label{fig:logit_normal_mape_experiments}
\end{figure}

Define the function $g(x,y)$ to be the joint density of $X = 1/(1 + e^{-G_1})$ and $Y = 1/(1 + e^{-G_2})$, such that $G=(G_1, G_2)$ and $G \sim \operatorname{Normal}(\mu, \Sigma)$, where $\mu \in \R^2$ and $\Sigma \in \R^{2\times 2}$ is a covariance matrix\footnote{Since the moments of $X$ and $Y$ are not closed-form, we obtain them through Monte Carlo.}.
For this experiment, we set two representative sets of parameters: $\mu = (0,0)$ and $\Sigma = [[1, 0.1], [0.1,1]]$ and $\mu = -(1,1)$ and $\Sigma = [[2.25,-1.2], [-1.2,1]]$, and simulate $n=50$ samples to get estimates for $\parameter$.
Then, we use Monte Carlo to estimate bias, MAPE and MSE comparing the true moments and the estimated through $\hat\parameter$ with $1,000$ simulations.
The results are summarised in Tables~\ref{tab:alpha-experiment4} and~\ref{tab:alpha-experiment5}.
In the first experiment, the true moments are, approximately $\ev[X] = \ev[Y] = 0.5, \var(X) = \var(Y) = 0.0433$ and $\cor(X, Y) = 0.098$. 
With these values, the MM1 estimate is very precise since means and variances obey the relation given in~\autoref{prop:solution-to-system-bivariate-beta}. 
The other estimators are very similar.
Despite that, the MAPE value is greater than 100\% for the correlation estimate as we can observe in~\autoref{fig:logit_normal_mape_experiments}.
This happens because sample correlation, which we use as an estimator for the correlation between the random variables, is not a good estimator of $\rho$.
In~\autoref{fig:sample_distribution_rho}, we observe that the sample distribution of $P_n$ is too wide.
In particular, we estimate that $\pr(P_n \not\in [0,0.2]) \approx 0.5$, that is 
\[
\pr\left(\bigg|\frac{P_n - \rho}{\rho}\bigg| > 1\right) \approx 0.5,
\]
which explains the higher value in the correlation, considering that the method of moments approximates the empirical correlation, rather than the true correlation, by construction.

\begin{figure}[!htbp]
    \centering
    \includegraphics[width=0.7\textwidth]{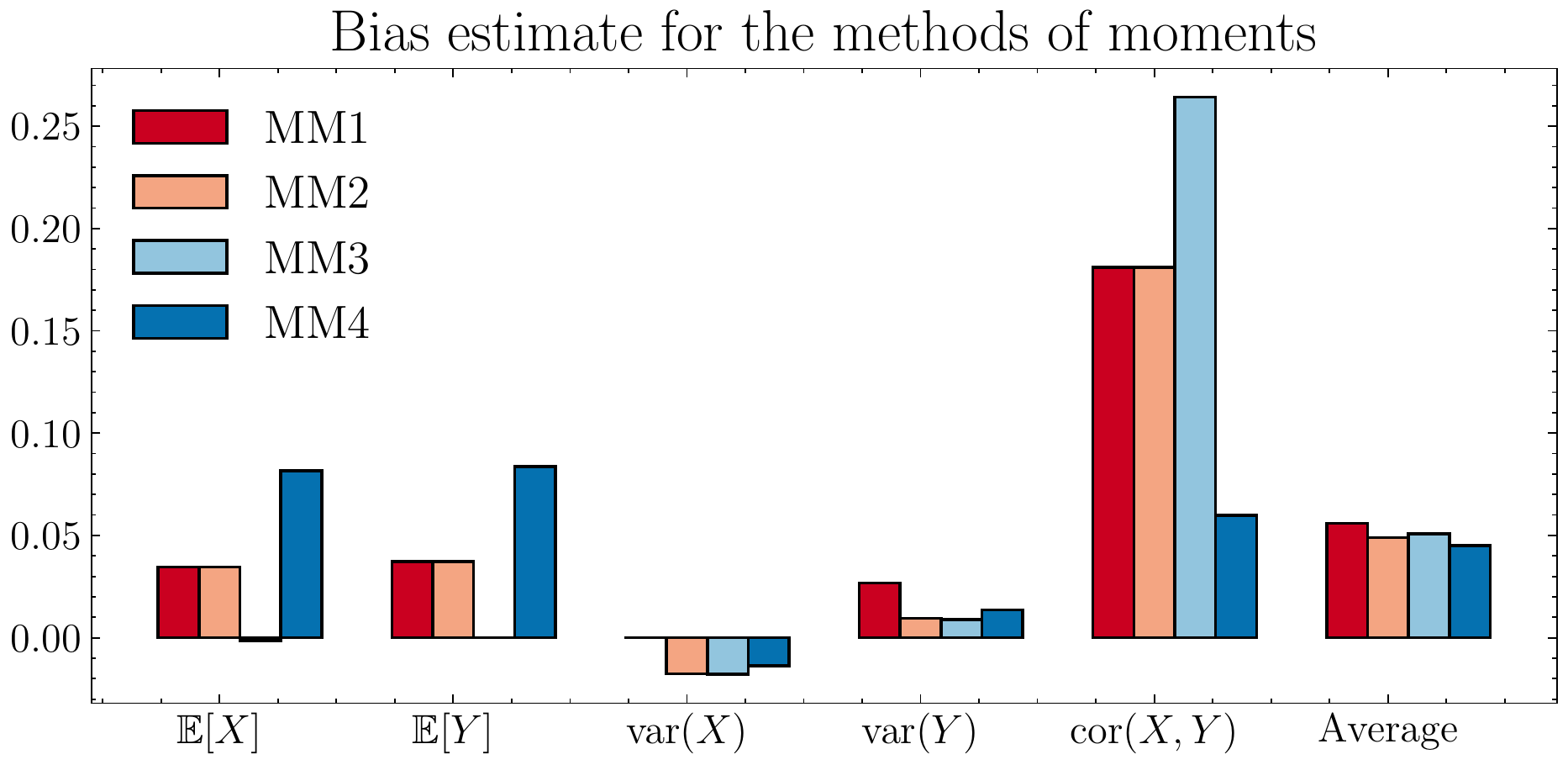}
    \caption{\textbf{Bias for the estimates of each moment:}
    for each moment, we calculate the bias between the true value and the corresponding one from the bivariate beta with the estimated $\hat{\boldsymbol\alpha}$ when $\mu=[-1,-1]$ and $\Sigma = [[2.25,-1.2],[-1.2,1]]$.}\label{fig:bias_estimate_moments_logit_normal}
\end{figure}

The second experiment has the true moments $\ev[X] = 0.33$, $\ev[Y] = 0.3$, $\var(X) = 0.062$, $\var(Y) = 0.033$ and $\cor(X,Y) = -0.73$.
These values do not yield a well-defined bivariate beta because the solution in~\eqref{eq:system-solution} does not belong to the parameter space.
Therefore, each method proposes approximations based on how it was defined.
In~\autoref{fig:logit_normal_mape_experiments} we notice that MM1 has a larger error in the second variance since it ignores its value. 
On the other side, MM4 compensates by increasing the error in the means and the first variance.
In real applications, one should think about the preferences for each situation: for instance, if one cares about getting the marginal means very precisely but does not care much for the correlation, MM3 poses a good alternative.
The difference between the methods is also seen through the bias, as presented in~\autoref{fig:bias_estimate_moments_logit_normal}.
For instance, MM3 solves the equation exactly for the means, yielding zero bias.
Despite MM1 and MM2 also solving for these quantities, since the solution is negative, the negative values are replaced by zero. 
The Bayesian estimates are similar to the method of moments for these analyses.

\begin{figure}
    \centering
    \includegraphics[width=\textwidth]{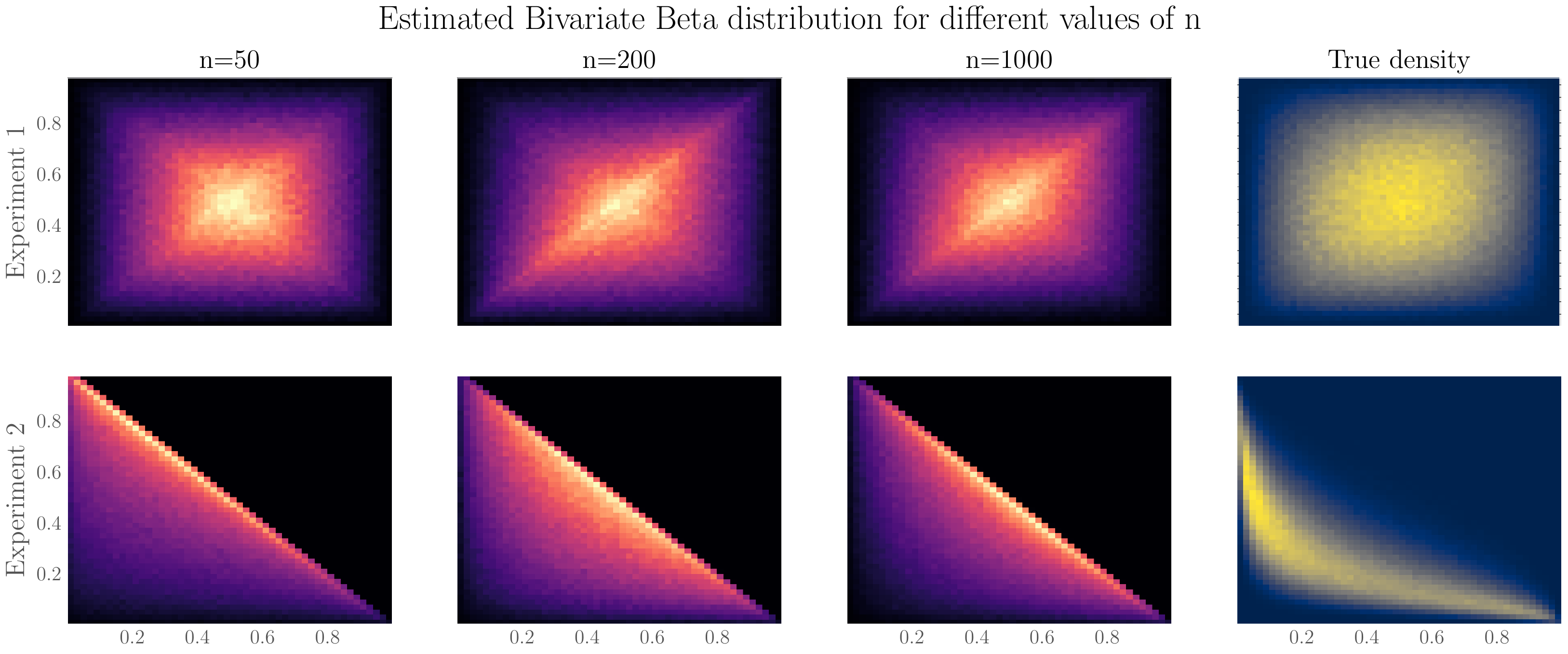}
    \caption{{\bf Comparing the contour plots} of the bivariate beta density with parameter $\hat\parameter$ and the true density of $g$ for both experiments in the misspecified case.
    MM4 was used as the estimator based on the different number of samples: $n=50, 200$ and $1,000$.
    }\label{fig:estimated_distributions_against_true}
\end{figure}

Finally, we turn our attention to the whole density we want to approximate --- see~\autoref{fig:estimated_distributions_against_true}.
In the first experiment, the estimated moments are close to the true ones, especially when $n$ is large.
The densities are similar, but $g$ seems more dispersed.
If we compare the marginal kurtosis of each distribution, we notice that the bivariate beta model produced a value of $-3$ for each marginal against $-0.85$ of the true distribution.
Therefore, even in the well-specified case, this distribution may not represent other desired aspects.
In the same fashion, the bivariate beta in the second experiment represents the mode of the distribution poorly, driven by a poor method of moments estimate.
The estimated distribution does however give small probabilities to regions where the true probability is also low.

\subsection{Diagnostics}\label{sec:diagnostics}

As we have discussed in~\autoref{sec:misspecification}, the bivariate beta distribution does not always well approximate the unknown distribution of the data $\dd$.
This is not a problem with the distribution itself but reveals its inflexibility.
In this section, we propose a systematic way of diagnosing the compatibility of a given data set with the bivariate beta distribution.
% Let $(x_1, y_1), \dots, (x_n, y_n)$ be observations from a random sample from a distribution with support over $[0,1]^2$.

\subsubsection{Asymptotic diagnostics}

\begin{figure}[!ht]
    \centering 
\includegraphics[width=\textwidth]{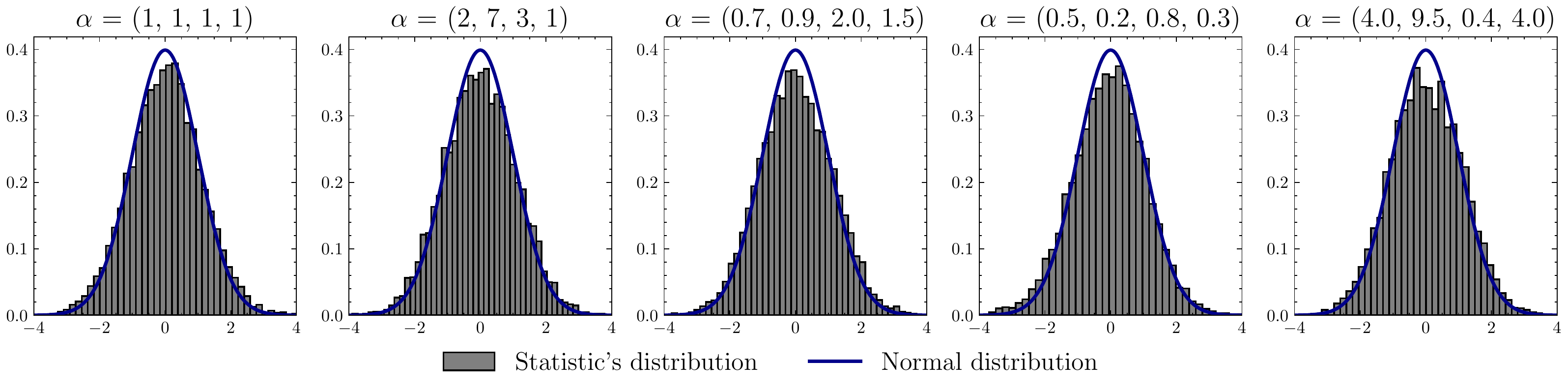}
    \caption{{\bf Comparison between the probability density function} of the statistic $\sqrt{n} S/\hat\sigma$ and of the standard normal distribution with $n=30$ and five specifications of the parameter $\parameter$.}\label{fig:distribution_statistic_n30}
\end{figure}

For the first diagnostic, we consider the relation established by~\autoref{prop:solution-to-system-bivariate-beta},
\[
\frac{m_1(1-m_1)}{v_1} = \frac{m_2(1-m_2)}{v_2}.
\]
This is a necessary condition for having a bivariate beta with marginal means $m_1$ and $m_2$ and marginal variances $v_1$ and $v_2$.
Define the function
\[
g(x_1,x_2,x_3,x_4) = x_1(1-x_1)x_4 - x_2(1-x_3)x_3
\]
and the statistic 
\[
G_n = g(\bar{X}_n, S_{X,n}^2, \bar{Y}_n, S_{Y,n}^2).
\]
If data comes from a bivariate beta distribution, by the consistency property of the method of moments, $G_n$ converges to $0$ in probability as $n$ tends to infinity.
Moreover, as proved in~\autoref{appendix:asymptotic-distribution},
\[
\sqrt{n}G_n \overset{d}{\to} N\left(0, \nabla g{(m_1, v_1, m_2, v_2)}^T \Sigma \nabla g(m_1, v_1, m_2, v_2)\right).
\]
and, consequently, 
\[
\sqrt{n} \frac{G_n}{\hat\sigma_n} \to N(0,1),
\]
where $\hat\sigma^2_n$ is a consistent estimator for $\nabla g{(m_1, v_1, m_2, v_2)}^T \Sigma \nabla g(m_1, v_1, m_2, v_2)$. 
See~\autoref{fig:distribution_statistic_n30} for numerical examples with $n=30$ comparing the distribution of $\sqrt{n} G_n/\hat\sigma_n$, estimated through Monte Carlo, and the standard normal distribution.
The testing procedure tests the null hypotheses that the data comes from a bivariate beta distribution and $g(m_1, v_1, m_2, v_2) = 0$, against the alternative that $g(m_1, v_1, m_2, v_2) \neq 0$.
If $\sqrt{n} G_n/\hat\sigma_n = s$ is observed for a random sample of size $n$, the $p$-value is $p = 2\Phi(-|s|)$, where $\Phi$ is the standard normal cumulative distribution function (CDF). 

\autoref{fig:comparing_pvalue_distributions} compares the distribution of the $p$-value under the null and alternative hypothesis fixing a bivariate beta distribution as control and four different independent beta specifications to see how the statistics behave.

The second diagnostic we introduce captures the correlation aspect of~\autoref{prop:solution-to-system-bivariate-beta}.
If the data comes from a bivariate beta, its marginal means and variances and correlation must satisfy the relation in~\eqref{eq:system-solution} with a positive solution.
Suppose that $S_{X,n}^2 < \bar{X}_n(1-\bar{X}_n)$ and let $\hat{\parameter}$ be the solution given by the system in~\eqref{eq:system-solution}. 
Define $\hat{\beta}_i = \hat{\alpha}_i/\bar{\alpha}$ and the summary statistic 
\[
M = \min\{\hat{\beta}_1, \dots, \hat{\beta}_4\}.
\]
Since $M > 0 \iff \hat{\alpha}_i > 0, i=1,\dots,4$, the proposed procedure rejects the null hypothesis that data comes from a bivariate beta distribution if $M \le c$ for some fixed $c$.
We observe that if $n$ goes to infinity, this test never falsely rejects the null hypothesis considering $c=0$ because the method of moments is consistent.
For a finite value of $n$, there is a chance that, if some $\alpha_i$ is small, the method of moments may yield a negative solution.

The distribution of $M$ is hard to derive, even asymptotically, because it is the minimum of correlated statistics.
So a non-parametric Bootstrap approach is appealing.
\autoref{fig:quantile_distribution_minimum} shows the distribution of the 5th-quantile of the distribution of $M$ for $n=50$ taking uniformly random values of $\parameter$ in ${[0,0.5]}^4$.
This region of small values is the main cause of negative values in the method of moments. 
Empirically, we notice that if we set $c=-0.05$, the probability of rejection under the null hypothesis is at most $0.05$. 
In summary, we reject the hypothesis that data comes from a bivariate beta distribution if $M \le -0.05$.
\autoref{fig:quantile_distribution_minimum} also shows the distribution of 1st and 10th quantiles, but there is not much difference. 

\subsubsection{Bayesian diagnostics}

\begin{figure}[!ht]
    \centering
    \includegraphics[width=\textwidth]{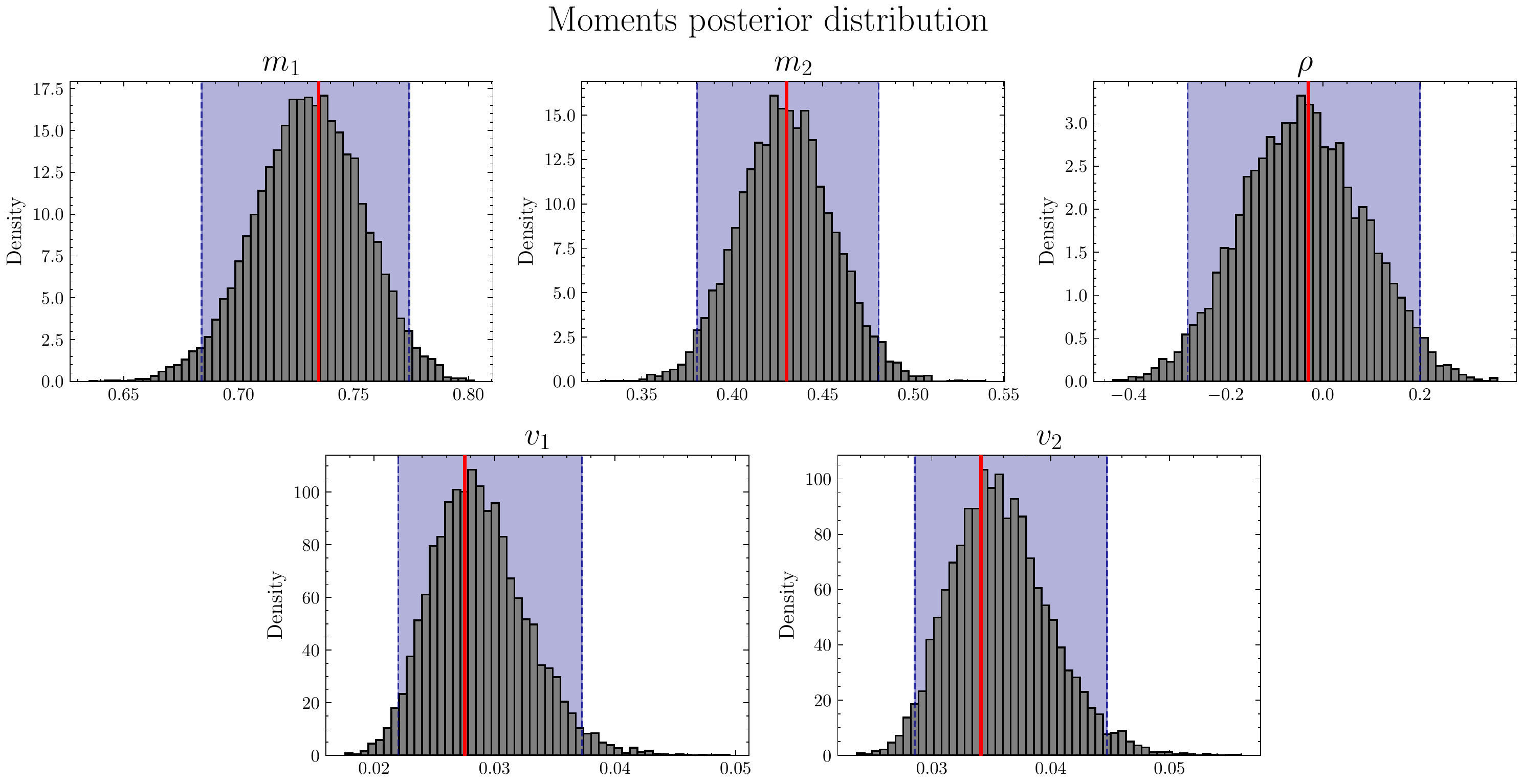}
    \caption{{\bf Posterior predictive checks:} Comparing the posterior distribution of the moments (histogram in grey) and the observed values (solid red line).
    We set $\parameter = (2,3,0.7,1)$ and $n=50$ to generate the data, with independent Gamma$(1,1)$ as priors.
    The blue-coloured region is the$95\%$ equal-tailed probability interval.}\label{fig:posterior_moments_distribution}
\end{figure}

For diagnosing the Bayes estimator, there are some tools we can use.
The posterior predictive checks (PPC) use the parameters' posterior samples to see induced variables, comparing them to the observed in the sample.
In our case, we can compare the observed moments $\hat{m}_1, \hat{m}_2, \hat{v}_1, \hat{v}_2$ and $\hat{\rho}$ to the calculated through the samples of the posterior $p(\parameter | \dd)$.
\autoref{fig:posterior_moments_distribution} shows an example of PPC.\@
We notice that the observed moments are in the interval between the 2.5th and 97.5th quartiles.
Prior information can improve these results.

\section{Discussion and conclusions}\label{sec:conclusion}

In this paper, we have dealt with a particular construction of the bivariate distribution~\citep{olkin2015constructions}, which strikes a balance between parameter-richness and tractability.
We leverage a detailed mathematical analysis of the moments to propose a class of moment-based estimators for the parameter of interest, $\parameter \in \mathbb{R}_+^4$.
We also exploit a latent variable construction to propose an efficient representation of the posterior distribution in order to obtain Bayesian estimates.
Finally, we have discussed the construction of simple yet powerful diagnostics to evaluate model fit, from both frequentist and Bayesian perspectives.

We find that no single method performs best under all situations, but that there is an inherent trade-off between statistical and computational performance: moments estimators that are based on solving fewer equations numerically tend to have lower runtime but also perform worse in terms of bias or MAPE.\@
In general, the bivariate beta provides a simple model which can be quickly fitted to data, but our experiments in the misspecified case show that it can sometimes be a rather inflexible distribution and fail to correctly capture the correlation between the data coordinates, which is a major feature of any bivariate model.

Future research will focus on testing a plethora of bivariate models, including the one analysed here on real data.
We have a special interest in sensitivity/specificity data, which are useful when analysing imperfect test data under a Bayesian paradigm~\citep{gelman2020bayesian}.

In summary, we hope to have provided applied researchers with the tools to quickly fit the bivariate beta model to data and diagnose problems, while  at the same time presenting the methods community with a modern discussion of a complete suite of tools to guide model development and assessment in the case of correlated proportions.
    
\section*{Acknowledgements}
We thank Rodrigo Targino and Eduardo Mendes for their insightful discussions.
The first author thanks the financial support from the School of Applied Mathematics (FGV EMAp).

% \newpage
\bibliographystyle{apalike}
\bibliography{bivariate_beta} 

\appendix
\makeatletter 
\renewcommand\thefigure{\thesection.\arabic{figure}}
\renewcommand\thetable{\thesection.\arabic{table}}
\makeatother

\section{Proofs}\label{sec:appendix-proof}

\printProofs{}

\subsection{Code for the solution of the three-equation system}\label{sec:code-solution-three-equations}

\begin{lstlisting}[language=Python]
import sympy as sp

if __name__ == '__main__':
    m1, m2, rho, alpha3, alpha4 = sp.symbols('m1 m2 rho alpha3 alpha4')
    alpha1 = (m1 + m2 - 1)/(1 - m1) * alpha3 + m2/(1 - m1) * alpha4
    alpha2 = (1 - m2)/(1 - m1) * alpha3 + (m1 - m2)/(1 - m1) * alpha4
    alpha_sum = sp.simplify(alpha1 + alpha2 + alpha3 + alpha4)
    expression = rho - (alpha1 * alpha4 - alpha2 * alpha3) 
    expression /= (alpha_sum**2 * sp.sqrt(m1 * m2 * (1-m1) * (1 - m2)))
    alpha3 = sp.simplify(sp.solve(expression, alpha3)[0])
    alpha1 = sp.simplify((m1 + m2 - 1)/(1 - m1) * alpha3 + m2/(1 - m1) * alpha4)
    alpha2 = sp.simplify((1 - m2)/(1 - m1) * alpha3 + (m1 - m2)/(1 - m1) * alpha4)
    print('alpha1 = {}'.format(alpha1))
    print('alpha2 = {}'.format(alpha2))
    print('alpha3 = {}'.format(alpha3))
\end{lstlisting}

\if{ 
\input{archive/mle-archived.tex}
}\fi

\newpage
\section{Additional figures}

\begin{figure}[!ht]
  \centering
  \includegraphics[width=14cm]{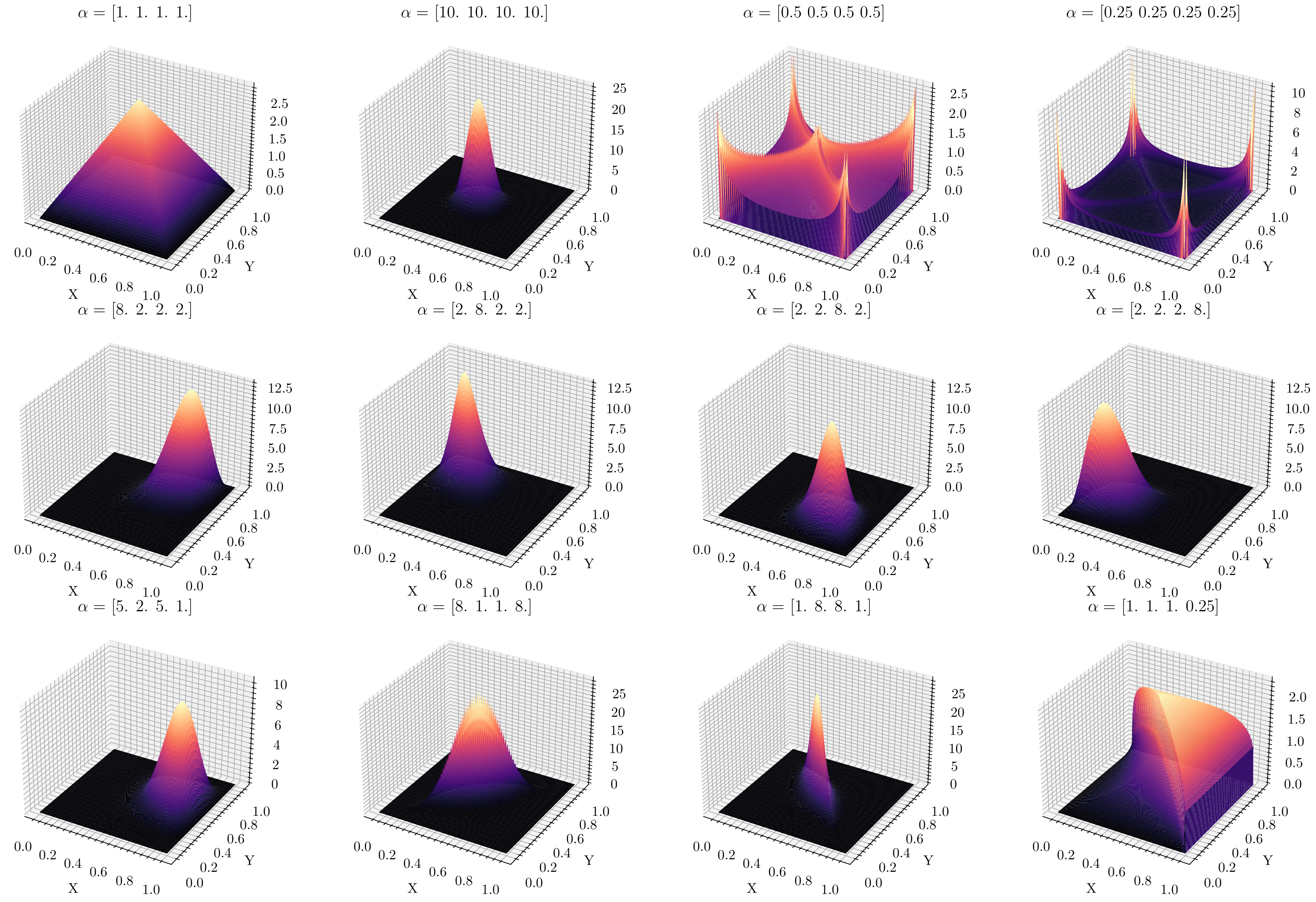}
  \caption{{\bf Joint density of the variables $X$ and $Y$:} The bivariate density for different choices of $\boldsymbol{\alpha}$. The four plots in the first row are   symmetric with respect to the mode and have no correlation between the variables $X$ and $Y$.}\label{fig:beta-bivariate}
\end{figure}

\begin{figure}
    \centering
    \includegraphics[width=0.6\textwidth]{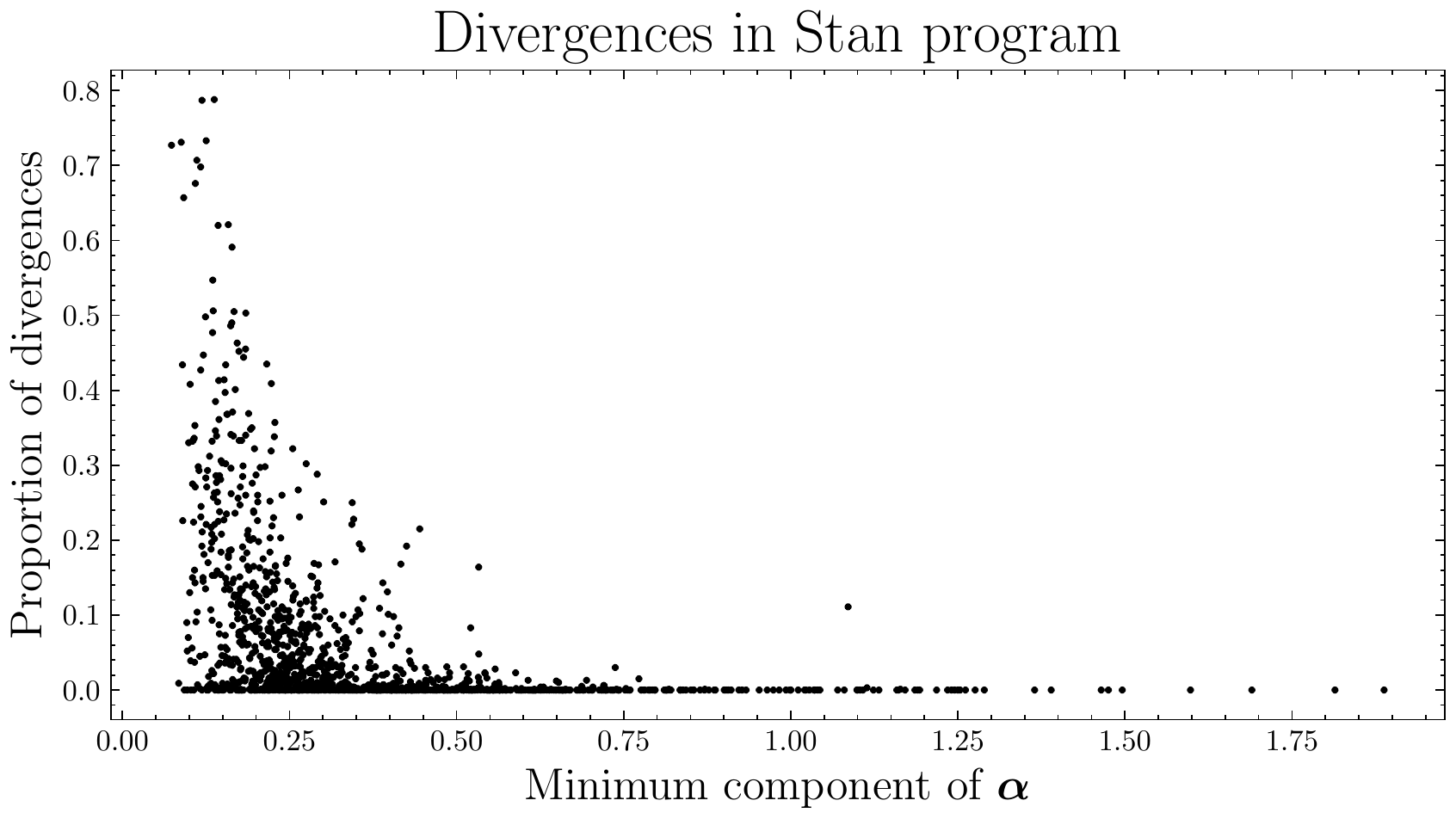}
    \caption{{\bf Divergences in Stan program:} The proportion of divergences in Stan program for each sampled $\tilde{\parameter}$ from simulation-based calibration.
    If some $\alpha_i < 0.5$, divergences happen more than we would like, biasing the results.}\label{fig:sbc_divergences}
\end{figure}

\begin{figure}
    \centering
    \includegraphics[width=0.8\textwidth]{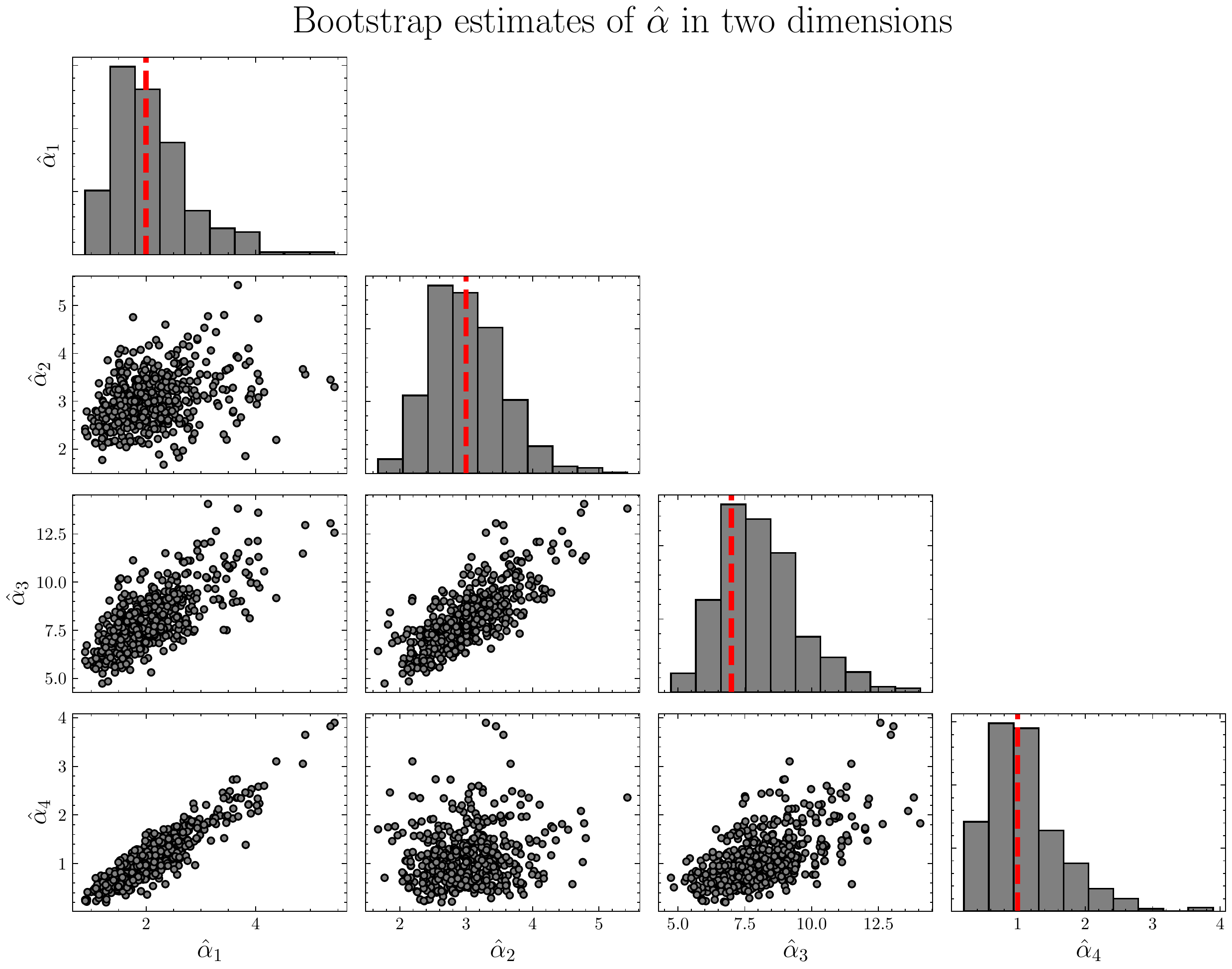}
    \caption{{\bf Bootstrap estimates in two dimensions:} Pairwise estimates of each component of $\parameter$ given by bootstrap method with $B=500$.
    The lines in red show the true value of $\parameter = (2,3,7,1)$ that generated the original dataset of size $n=50$.}\label{fig:bootstrap_samples_2d}
\end{figure}

\begin{figure}[!htbp]
    \centering
    \includegraphics[width=0.6\textwidth]{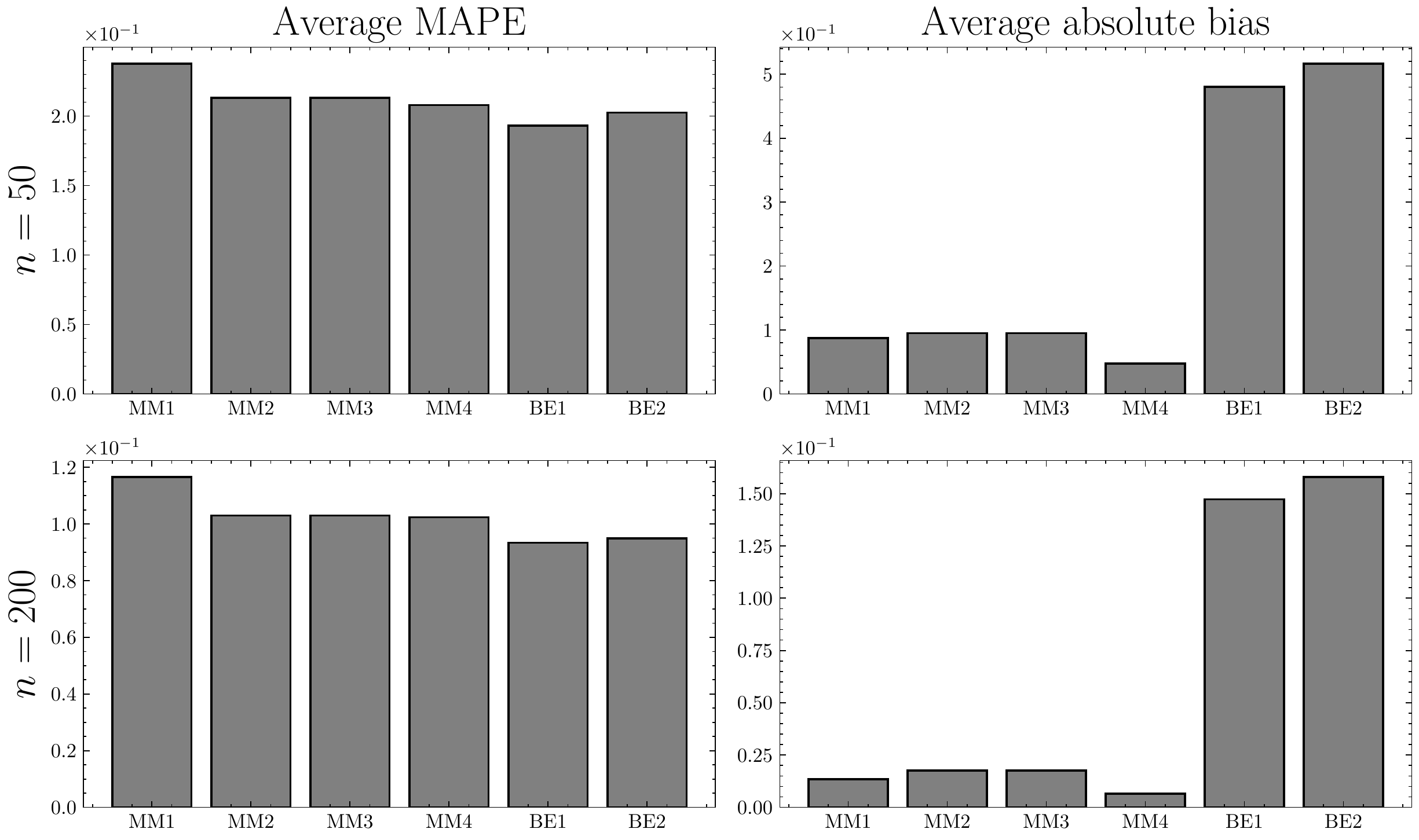}
    \caption{{\bf Estimation performance} when the true value is $\boldsymbol{\alpha} = (2,7,3,1)$.
    The average MAPE is the mean of the MAPEs for each estimate $\hat\alpha_i$, while the average absolute bias considers the mean of the absolute values of the bias of each estimate.
    }\label{fig:comparing_methods_mape_bias003}
\end{figure}

\begin{figure}
    \centering
    \includegraphics[width=\textwidth]{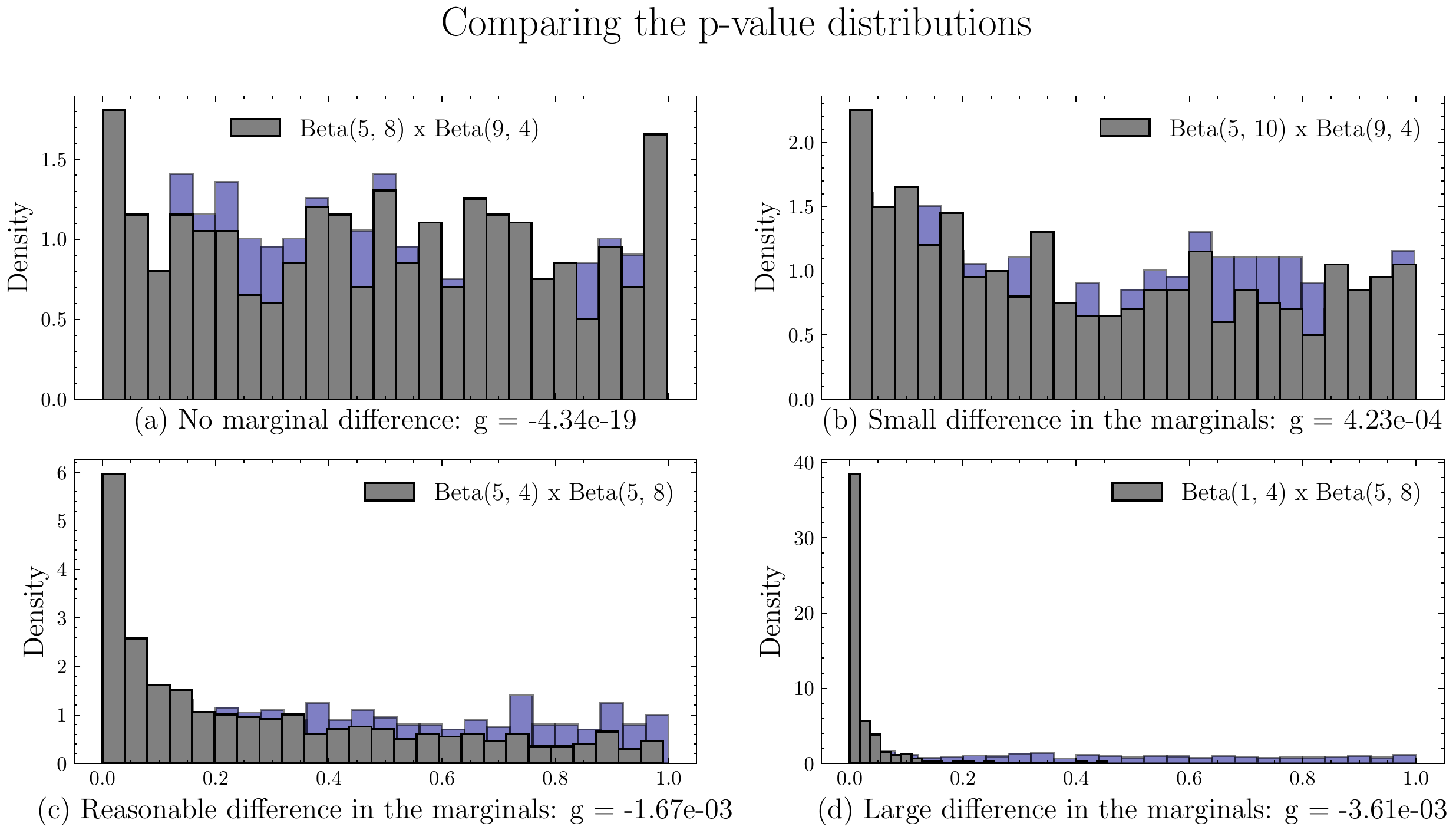}
    \caption{{\bf Comparing the p-value distributions:} we compare the distribution of the p-value under the hypothesis that data came from a bivariate beta distribution with parameter $(2,3,7,1)$ in blue, against
    two independent beta distributions with specified parameters.}\label{fig:comparing_pvalue_distributions}
\end{figure}

\begin{figure}
    \centering
    \includegraphics[width=\textwidth]{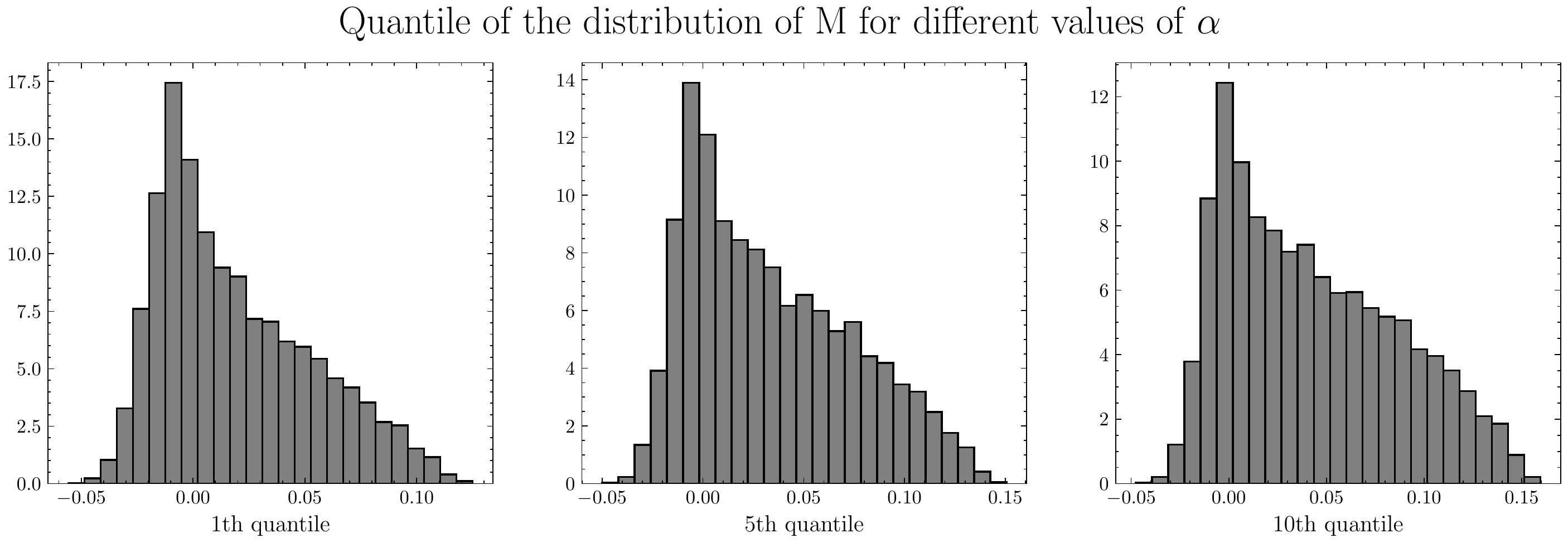}
    \caption{{\bf Quantile distribution:} values of the 1st, 5th and 10th quantiles of the distribution of the statistic $M$ for different specifications of $\parameter$, which is chosen uniformly random in ${[0,0.5]}^4$.}\label{fig:quantile_distribution_minimum}
\end{figure}

\newpage
\section{Additional tables}\label{appendix:tables}

\begin{table}[htbp]
\centering
\begin{tabular}{c|c|cccc|cccc}
                     &            & \multicolumn{4}{c|}{$n=50$}                       & \multicolumn{4}{c}{$n=200$}                      \\ \hline
Method               & Evaluation & $\alpha_1$ & $\alpha_2$ & $\alpha_3$ & $\alpha_4$ & $\alpha_1$ & $\alpha_2$ & $\alpha_3$ & $\alpha_4$ \\ \hline
\multirow{4}{*}{MM1}& Bias ($10^{-2}$)  & 1.79 & 4.08 & 4.01 & 2.43 & 0.23 & 0.33 & -0.29 & 0.31 \\ 
& MSE ($10^{-1}$)  & 0.78 & 0.78 & 0.76 & 0.84 & 0.17 & 0.17 & 0.17 & 0.18 \\ 
& MAPE ($10^{-1}$)  & 2.23 & 2.18 & 2.11 & 2.25 & 1.04 & 1.05 & 1.01 & 1.04 \\ 
& Coverage (\%)  & 94.7 & 93.9 & 95.1 & 93.0 & 94.4 & 93.5 & 94.1 & 93.2 \\ 
\hline 
\multirow{4}{*}{MM2}& Bias ($10^{-2}$)  & 2.12 & 4.32 & 4.31 & 2.68 & 0.41 & 0.52 & -0.09 & 0.46 \\ 
& MSE ($10^{-1}$)  & 0.63 & 0.6 & 0.59 & 0.66 & 0.13 & 0.13 & 0.13 & 0.13 \\ 
& MAPE ($10^{-1}$)  & 2.01 & 1.91 & 1.87 & 2.02 & 0.9 & 0.91 & 0.9 & 0.9 \\ 
& Coverage (\%)  & 93.0 & 93.8 & 94.3 & 92.4 & 94.4 & 94.3 & 95.3 & 95.5 \\ 
\hline 
\multirow{4}{*}{MM3}& Bias ($10^{-2}$)  & 2.12 & 4.32 & 4.31 & 2.68 & 0.41 & 0.52 & -0.09 & 0.46 \\ 
& MSE ($10^{-1}$)  & 0.63 & 0.6 & 0.59 & 0.66 & 0.13 & 0.13 & 0.13 & 0.13 \\ 
& MAPE ($10^{-1}$)  & 2.01 & 1.91 & 1.87 & 2.02 & 0.9 & 0.91 & 0.9 & 0.9 \\ 
& Coverage (\%)  & 93.8 & 94.0 & 94.4 & 92.8 & 95.2 & 94.4 & 94.7 & 95.5 \\ 
\hline 
\multirow{4}{*}{MM4}& Bias ($10^{-2}$)  & 0.88 & 3.06 & 3.04 & 1.44 & 0.11 & 0.22 & -0.39 & 0.16 \\ 
& MSE ($10^{-1}$)  & 0.61 & 0.57 & 0.56 & 0.63 & 0.13 & 0.13 & 0.13 & 0.13 \\ 
& MAPE ($10^{-1}$)  & 1.98 & 1.87 & 1.82 & 1.98 & 0.89 & 0.91 & 0.9 & 0.9 \\ 
& Coverage (\%)  & 93.3 & 93.8 & 94.8 & 93.1 & 95.1 & 95.1 & 95.1 & 95.2 \\ 
\hline 
\multirow{4}{*}{BE1}& Bias ($10^{-2}$)  & 1.91 & 3.71 & 3.65 & 2.27 & 0.53 & 0.44 & -0.14 & 0.59 \\ 
& MSE ($10^{-1}$)  & 0.45 & 0.44 & 0.42 & 0.46 & 0.11 & 0.11 & 0.11 & 0.11 \\ 
& MAPE ($10^{-1}$)  & 1.67 & 1.65 & 1.58 & 1.7 & 0.81 & 0.84 & 0.83 & 0.83 \\ 
& Coverage (\%)  & 94.6 & 95.9 & 96.1 & 95.2 & 95.3 & 94.4 & 94.8 & 95.5 \\ 
\hline 
\multirow{4}{*}{BE2}& Bias ($10^{-2}$)  & 0.7 & 2.51 & 2.45 & 1.07 & 0.24 & 0.14 & -0.43 & 0.3 \\ 
& MSE ($10^{-1}$)  & 0.44 & 0.44 & 0.41 & 0.46 & 0.11 & 0.11 & 0.11 & 0.11 \\ 
& MAPE ($10^{-1}$)  & 1.67 & 1.64 & 1.58 & 1.7 & 0.82 & 0.84 & 0.83 & 0.83 \\ 
& Coverage (\%)  & 94.6 & 95.9 & 96.1 & 95.2 & 95.3 & 94.4 & 94.8 & 95.5 \\ 
\end{tabular}
\caption{\label{tab:alpha-experiment1}Estimate of bias, MSE, MAPE and Coverage for each of the six methods when the true value of $\alpha$ of the generative process is $\alpha = (1,1,1,1)$ and the number of samples is $n=50$ or $n=200$. 
The estimates are calculated using Monte Carlo with $1,000$ iterations, as described in \autoref{sec:recovering-bivariate-beta}.}
\end{table}

\begin{table}[htbp]
\centering
\begin{tabular}{c|c|cccc|cccc}
                     &            & \multicolumn{4}{c|}{$n=50$}                       & \multicolumn{4}{c}{$n=200$}                      \\ \hline
Method               & Evaluation & $\alpha_1$ & $\alpha_2$ & $\alpha_3$ & $\alpha_4$ & $\alpha_1$ & $\alpha_2$ & $\alpha_3$ & $\alpha_4$ \\ \hline
\multirow{4}{*}{MM1}& Bias ($10^{-2}$)  & 6.44 & 17.56 & 6.58 & 4.26 & 1.48 & 1.81 & 0.61 & 1.5 \\ 
& MSE ($10^{-1}$)  & 4.63 & 21.89 & 4.15 & 2.45 & 1.0 & 5.04 & 0.95 & 0.53 \\ 
& MAPE ($10^{-1}$)  & 2.51 & 1.61 & 1.69 & 3.7 & 1.24 & 0.81 & 0.82 & 1.8 \\ 
& Coverage (\%)  & 95.5 & 94.6 & 95.0 & 94.4 & 94.9 & 95.0 & 95.0 & 93.9 \\ 
\hline 
\multirow{4}{*}{MM2}& Bias ($10^{-2}$)  & 6.62 & 19.42 & 7.38 & 4.6 & 1.7 & 2.64 & 1.05 & 1.64 \\ 
& MSE ($10^{-1}$)  & 3.53 & 13.71 & 2.72 & 2.17 & 0.81 & 2.9 & 0.61 & 0.49 \\ 
& MAPE ($10^{-1}$)  & 2.26 & 1.3 & 1.36 & 3.6 & 1.12 & 0.61 & 0.66 & 1.73 \\ 
& Coverage (\%)  & 94.1 & 93.6 & 93.9 & 94.2 & 94.4 & 94.9 & 95.8 & 93.9 \\ 
\hline 
\multirow{4}{*}{MM3}& Bias ($10^{-2}$)  & 6.62 & 19.42 & 7.38 & 4.6 & 1.7 & 2.64 & 1.05 & 1.64 \\ 
& MSE ($10^{-1}$)  & 3.53 & 13.71 & 2.72 & 2.17 & 0.81 & 2.9 & 0.61 & 0.49 \\ 
& MAPE ($10^{-1}$)  & 2.26 & 1.3 & 1.36 & 3.6 & 1.12 & 0.61 & 0.66 & 1.73 \\ 
& Coverage (\%)  & 95.2 & 94.3 & 95.3 & 94.7 & 94.3 & 95.0 & 96.1 & 93.9 \\ 
\hline 
\multirow{4}{*}{MM4}& Bias ($10^{-2}$)  & 3.56 & 9.23 & 3.11 & 3.03 & 1.0 & 0.26 & 0.04 & 1.29 \\ 
& MSE ($10^{-1}$)  & 3.27 & 12.52 & 2.55 & 2.05 & 0.8 & 2.85 & 0.61 & 0.48 \\ 
& MAPE ($10^{-1}$)  & 2.2 & 1.25 & 1.33 & 3.53 & 1.12 & 0.6 & 0.66 & 1.72 \\ 
& Coverage (\%)  & 95.6 & 95.4 & 94.9 & 94.0 & 94.1 & 96.0 & 96.3 & 94.4 \\ 
\hline 
\multirow{4}{*}{BE1}& Bias ($10^{-2}$)  & -26.35 & -111.25 & -44.04 & -10.55 & -8.27 & -34.25 & -13.3 & -3.12 \\ 
& MSE ($10^{-1}$)  & 2.02 & 17.48 & 3.21 & 1.03 & 0.57 & 3.35 & 0.64 & 0.32 \\ 
& MAPE ($10^{-1}$)  & 1.87 & 1.66 & 1.61 & 2.58 & 0.95 & 0.68 & 0.68 & 1.42 \\ 
& Coverage (\%)  & 89.2 & 74.6 & 81.3 & 91.1 & 92.8 & 88.7 & 90.2 & 92.8 \\ 
\hline 
\multirow{4}{*}{BE2}& Bias ($10^{-2}$)  & -30.66 & -115.58 & -45.6 & -14.83 & -9.53 & -35.49 & -13.79 & -4.36 \\ 
& MSE ($10^{-1}$)  & 2.24 & 18.4 & 3.32 & 1.12 & 0.59 & 3.43 & 0.65 & 0.33 \\ 
& MAPE ($10^{-1}$)  & 2.0 & 1.71 & 1.65 & 2.73 & 0.97 & 0.69 & 0.69 & 1.44 \\ 
& Coverage (\%)  & 89.2 & 74.6 & 81.3 & 91.1 & 92.8 & 88.7 & 90.2 & 92.8 \\ 
\end{tabular}
\caption{\label{tab:alpha-experiment2}Estimate of bias, MSE, MAE and Coverage for each of the six methods when the true value of $\alpha$ of the generative process is $\alpha = (2,7,3,1)$ and the number of samples is $n=50$ and $n=200$. 
The estimates are calculated using Monte Carlo using $1,000$ iterations, as described in \autoref{sec:recovering-bivariate-beta}.}
\end{table}

\begin{table}[htbp]
\centering
\begin{tabular}{c|c|cccc|cccc}
                     &            & \multicolumn{4}{c|}{$n=50$}                       & \multicolumn{4}{c}{$n=1000$}                      \\ \hline
Method               & Evaluation & $\alpha_1$ & $\alpha_2$ & $\alpha_3$ & $\alpha_4$ & $\alpha_1$ & $\alpha_2$ & $\alpha_3$ & $\alpha_4$ \\ \hline
\multirow{4}{*}{MM1}& Bias ($10^{-2}$)  & 1.56 & 1.34 & 5.19 & 3.6 & 0.42 & 0.54 & 1.5 & 1.38 \\ 
& MSE ($10^{-1}$)  & 0.63 & 0.69 & 2.5 & 1.67 & 0.16 & 0.17 & 0.52 & 0.4 \\ 
& MAPE ($10^{-1}$)  & 2.73 & 2.29 & 1.87 & 2.05 & 1.42 & 1.15 & 0.9 & 1.04 \\ 
& Coverage (\%)  & 93.96 & 94.31 & 94.66 & 95.35 & 94.6 & 94.5 & 96.4 & 94.4 \\ 
\hline 
\multirow{4}{*}{MM2}& Bias ($10^{-2}$)  & 2.12 & 1.77 & 5.57 & 4.29 & 0.43 & 0.55 & 1.39 & 1.33 \\ 
& MSE ($10^{-1}$)  & 0.55 & 0.55 & 1.48 & 1.21 & 0.14 & 0.14 & 0.32 & 0.29 \\ 
& MAPE ($10^{-1}$)  & 2.66 & 2.09 & 1.5 & 1.81 & 1.34 & 1.05 & 0.7 & 0.9 \\ 
& Coverage (\%)  & 94.03 & 95.14 & 94.52 & 94.38 & 92.7 & 93.1 & 95.8 & 94.8 \\ 
\hline 
\multirow{4}{*}{MM3}& Bias ($10^{-2}$)  & 2.12 & 1.77 & 5.57 & 4.29 & 0.43 & 0.55 & 1.39 & 1.33 \\ 
& MSE ($10^{-1}$)  & 0.55 & 0.55 & 1.48 & 1.21 & 0.14 & 0.14 & 0.32 & 0.29 \\ 
& MAPE ($10^{-1}$)  & 2.66 & 2.09 & 1.5 & 1.81 & 1.34 & 1.05 & 0.7 & 0.9 \\ 
& Coverage (\%)  & 93.75 & 94.73 & 95.07 & 93.2 & 93.0 & 93.9 & 94.9 & 95.0 \\ 
\hline 
\multirow{4}{*}{MM4}& Bias ($10^{-2}$)  & 1.11 & 0.48 & 2.57 & 2.05 & 0.16 & 0.2 & 0.59 & 0.73 \\ 
& MSE ($10^{-1}$)  & 0.53 & 0.53 & 1.37 & 1.14 & 0.14 & 0.14 & 0.31 & 0.29 \\ 
& MAPE ($10^{-1}$)  & 2.63 & 2.07 & 1.47 & 1.77 & 1.33 & 1.06 & 0.7 & 0.89 \\ 
& Coverage (\%)  & 94.73 & 94.45 & 95.49 & 93.62 & 93.3 & 94.0 & 95.8 & 94.4 \\ 
\hline 
\multirow{4}{*}{BE1}& Bias ($10^{-2}$)  & 2.31 & -2.46 & -4.37 & 0.73 & 0.65 & -0.66 & -1.3 & 0.63 \\ 
& MSE ($10^{-1}$)  & 0.34 & 0.39 & 1.05 & 0.77 & 0.11 & 0.12 & 0.28 & 0.23 \\ 
& MAPE ($10^{-1}$)  & 2.1 & 1.78 & 1.31 & 1.46 & 1.19 & 0.97 & 0.66 & 0.79 \\ 
& Coverage (\%)  & 95.84 & 95.07 & 94.86 & 95.14 & 94.7 & 94.1 & 94.8 & 95.2 \\ 
\hline 
\multirow{4}{*}{BE2}& Bias ($10^{-2}$)  & 1.04 & -3.3 & -6.07 & -0.96 & 0.31 & -0.83 & -1.71 & 0.19 \\ 
& MSE ($10^{-1}$)  & 0.35 & 0.4 & 1.06 & 0.77 & 0.11 & 0.12 & 0.28 & 0.23 \\ 
& MAPE ($10^{-1}$)  & 2.12 & 1.8 & 1.31 & 1.47 & 1.19 & 0.97 & 0.67 & 0.79 \\ 
& Coverage (\%)  & 95.84 & 95.07 & 94.86 & 95.14 & 94.7 & 94.1 & 94.8 & 95.2
\end{tabular}
\caption{\label{tab:alpha-experiment3}Estimate of bias, MSE, MAE and Coverage for each of the six methods when the true value of $\alpha$ of the generative process is $\alpha = (0.7, 0.9, 2, 1.5)$ for $n=50$ and $n=200$. 
The estimates are calculated using Monte Carlo using $1,000$ iterations, as described in \autoref{sec:recovering-bivariate-beta}.}
\end{table}

\begin{table}[htbp]
\centering
\begin{tabular}{c|c|ccccc}
Method & Evaluation & $\ev[X]$ & $\ev[Y]$ & $\var(X)$ & $\var(Y)$ & $\cov(X,Y)$ \\ 
\hline \multirow{3}{*}{MM1}& Bias ($10^{-2}$)  & -0.163 & -0.221 & 0.008 & 0.008 & 0.399 \\ 
& MSE ($10^{-2}$)  & 0.089 & 0.096 & 0.004 & 0.004 & 2.015 \\ 
& MAPE ($10^{-1}$)  & 0.475 & 0.494 & 1.202 & 1.196 & 11.85 \\ 
\hline \multirow{3}{*}{MM2}& Bias ($10^{-2}$)  & -0.163 & -0.221 & -0.049 & -0.049 & 0.399 \\ 
& MSE ($10^{-2}$)  & 0.089 & 0.096 & 0.002 & 0.002 & 2.015 \\ 
& MAPE ($10^{-1}$)  & 0.475 & 0.494 & 0.889 & 0.888 & 11.85 \\ 
\hline \multirow{3}{*}{MM3}& Bias ($10^{-2}$)  & -0.163 & -0.221 & -0.049 & -0.049 & 0.399 \\ 
& MSE ($10^{-2}$)  & 0.089 & 0.096 & 0.002 & 0.002 & 2.015 \\ 
& MAPE ($10^{-1}$)  & 0.475 & 0.494 & 0.889 & 0.888 & 11.85 \\ 
\hline \multirow{3}{*}{MM4}& Bias ($10^{-2}$)  & -0.163 & -0.221 & 0.001 & 0.001 & 0.399 \\ 
& MSE ($10^{-2}$)  & 0.089 & 0.096 & 0.002 & 0.002 & 2.015 \\ 
& MAPE ($10^{-1}$)  & 0.475 & 0.494 & 0.873 & 0.872 & 11.85 \\ 
\hline \multirow{3}{*}{BE1}& Bias ($10^{-2}$)  & -0.153 & -0.22 & -0.075 & -0.074 & -2.817 \\ 
& MSE ($10^{-2}$)  & 0.084 & 0.09 & 0.002 & 0.002 & 1.214 \\ 
& MAPE ($10^{-1}$)  & 0.458 & 0.481 & 0.732 & 0.732 & 9.076 \\ 
\hline \multirow{3}{*}{BE2}& Bias ($10^{-2}$)  & -0.154 & -0.223 & -0.036 & -0.036 & -2.69 \\ 
& MSE ($10^{-2}$)  & 0.085 & 0.091 & 0.002 & 0.002 & 1.252 \\ 
& MAPE ($10^{-1}$)  & 0.46 & 0.483 & 0.727 & 0.727 & 9.225 \\
\end{tabular}
\caption{\label{tab:alpha-experiment4}Estimate of bias, MSE and MAPE when $\mu = (0,0)$, $\sigma = [[1,0.1], [0.1,1]]$ and $n=50$, comparing the true moments and the estimated by the bivariate beta model.}
\end{table}

\begin{table}[htbp]
\centering
\begin{tabular}{c|c|ccccc}
Method & Evaluation & $\ev[X]$ & $\ev[Y]$ & $\var(X)$ & $\var(Y)$ & $\cov(X,Y)$ \\
\hline \multirow{3}{*}{MM1}& Bias ($10^{-2}$)  & 3.398 & 3.937 & 0.029 & 2.709 & 18.1 \\ 
& MSE ($10^{-2}$)  & 0.24 & 0.226 & 0.011 & 0.082 & 3.376 \\ 
& MAPE ($10^{-1}$)  & 1.218 & 1.353 & 1.354 & 8.13 & 2.494 \\ 
\hline \multirow{3}{*}{MM2}& Bias ($10^{-2}$)  & 3.398 & 3.937 & -1.743 & 0.994 & 18.1 \\ 
& MSE ($10^{-2}$)  & 0.24 & 0.226 & 0.035 & 0.015 & 3.376 \\ 
& MAPE ($10^{-1}$)  & 1.218 & 1.353 & 2.826 & 3.099 & 2.494 \\ 
\hline \multirow{3}{*}{MM3}& Bias ($10^{-2}$)  & -0.238 & 0.19 & -1.765 & 0.932 & 26.5 \\ 
& MSE ($10^{-2}$)  & 0.139 & 0.071 & 0.036 & 0.013 & 7.091 \\ 
& MAPE ($10^{-1}$)  & 0.903 & 0.694 & 2.863 & 2.948 & 3.652 \\ 
\hline \multirow{3}{*}{MM4}& Bias ($10^{-2}$)  & 8.151 & 8.596 & -1.358 & 1.396 & 5.875 \\ 
& MSE ($10^{-2}$)  & 0.789 & 0.831 & 0.024 & 0.024 & 0.568 \\ 
& MAPE ($10^{-1}$)  & 2.488 & 2.839 & 2.235 & 4.209 & 0.862 \\ 
\hline \multirow{3}{*}{BE1}& Bias ($10^{-2}$)  & -2.674 & 2.059 & -2.294 & 0.699 & 34.715 \\ 
& MSE ($10^{-2}$)  & 0.189 & 0.092 & 0.056 & 0.008 & 12.162 \\ 
& MAPE ($10^{-1}$)  & 1.088 & 0.813 & 3.717 & 2.203 & 4.784 \\ 
\hline \multirow{3}{*}{BE2}& Bias ($10^{-2}$)  & -2.843 & 1.894 & -2.275 & 0.721 & 33.919 \\ 
& MSE ($10^{-2}$)  & 0.198 & 0.086 & 0.055 & 0.008 & 11.616 \\ 
& MAPE ($10^{-1}$)  & 1.121 & 0.781 & 3.686 & 2.261 & 4.674 \\ 
\end{tabular}
\caption{\label{tab:alpha-experiment5}Estimate of bias, MSE and MAPE when $\mu = (-1,-1)$ and $\sigma = [[2.25,-1.2], [-1.2,1]]$ and $n=50$, comparing the true moments and the estimated by the bivariate beta model.}
\end{table}

\newpage
\section{Asymptotic distribution of S}\label{appendix:asymptotic-distribution}

By Proposition 2 in~\cite{arellano2021asymptotic}, 
\[
\sqrt{n}\begin{pmatrix}
    \bar{X}_n - m_1 \\
    S_{X,n}^2 - v_1 \\
    \bar{Y}_n - m_2 \\
    S_{Y,n}^2 - v_2
\end{pmatrix} \overset{d}{\rightarrow} N(0, \Sigma),
\]
for a covariance matrix $\Sigma$.
Notice that $\nabla g$ exists for all $(x_1, x_2, x_3, x_4) \in \R^4$ and it is continuous.
By multivariate Delta method, 
\[
\sqrt{n} G_n \overset{d}{\rightarrow} N\left(0, \nabla g{(m_1, v_1, m_2, v_2)}^T \Sigma \nabla g(m_1, v_1, m_2, v_2)\right),
\]
since $g(m_1, v_1, m_2, v_2) = 0$.

Using the paper's notation, for our application, $d=2, p=2$, $D_{11} = \bar{X}_n - m_1$, $D_{21} = \bar{Y}_n - m_2$, $S_{12} = (n-1) S_{X,n}^2/n$, $S_{22} = (n-1) S_{Y,n}^2/n$, $\kappa_{11} = 0$, $\kappa_{21} = 0$, $\kappa_{12} = v_1$ and $\kappa_{22} = v_2$.
By the Proposition, since $\ev[{(X-m_1)}^4]$ and $\ev[{(Y-m_2)}^4]$ are well-defined.
\[
\sqrt{n} \begin{pmatrix}
    D_{11} \\
    S_{12} \\
    D_{21} \\
    S_{22} 
\end{pmatrix} \overset{d}{\to} N_4\left(0, C\mathcal{K}C^T\right),
\]
where $C=I_4$ and $\mathcal{K} =$
\[
\begin{pmatrix}
    v_1 & \cov(X, {(X-m_1)}^2) & \rho\sqrt{v_1 v_2} & \cov(X, {(Y-m_2)}^2) \\
    \cov({(X-m_1)}^2, X) & \var{(X-m_1)}^2 & \cov({(X-m_1)}^2, Y) & \cov({(X-m_1)}^2, {(Y-m_2)}^2) \\
    \rho\sqrt{v_1 v_2}& \cov(Y, {(X-m_1)}^2) & v_2 & \cov(Y, {(Y-m_2)}^2) \\
    \cov({(Y-m_2)}^2, X) & \cov({(Y-m_2)}^2, {(X-m_1)}^2) & \cov({(Y-m_2)}^2, Y) & \var{(Y-m_2)}^2.
\end{pmatrix}
\]
Moreover, $S_{X,n}^2 \to S_{12}$ and $S_{Y,n}^2 \to S_{22}$.

By the continuity of $\nabla g$ and the consistency of the statistics $\bar{X}_n, \bar{Y}_n, S_{X,n}^2$ and $S_{Y,n}^2$ to $m_1, m_2, v_1$ and $v_2$, $\nabla g(\bar{X}_n, S_{X,n}^2, \bar{Y}_n, S_{Y,n}^2)$ is a consistent estimator for $\nabla g(m_1, v_1, m_2, v_2)$.
Therefore, we can calculate a consistent estimator for the variance of the limit distribution of $\hat{n} G_n$.
Let $\hat{\sigma}^2$ be this estimator. 
By Slutsky's theorem, we conclude that 
\[
\sqrt{n} \frac{G_n}{\hat\sigma} \to N(0,1).
\]

\section{Comments about integration}\label{sec:integration_comment}

The density of $(X,Y)$ is $f_{X,Y}(x,y)$ as in equation~\eqref{eq:dist-X-Y} and it can be undefined in sets of null Lebesgue measure in $\R^2$. 
These sets may be important when plotting on a grid, for instance.
This section illustrates one of these sets.

If $\alpha_i \ge 1, i = 1, \dots, 4$, the integral is clearly well defined for every $x,y \in [0,1]$. Let $0 < \alpha_2 = \alpha_3 = a \le 0.5$ and $x = y < 0.5$. 
Then
\begin{equation*}
  \begin{split}
    f_{X,Y}(x,y) &= \frac{1}{B(\boldsymbol{\alpha})}\int_{0}^x u_1^{\alpha_1-1}{(x-u_1)}^{a-1}{(x-u_1)}^{a-1}{(1-2x+u_1)}^{\alpha_4-1} \, du_1 \\
    &= \frac{1}{B(\boldsymbol{\alpha})}\int_{0}^{x/2} u_1^{\alpha_1-1}{(x-u_1)}^{2a-2}{(1-2x+u_1)}^{\alpha_4-1} \, du_1 + \\
    &~~~+ \frac{1}{B(\boldsymbol{\alpha})}\int_{x/2}^x u_1^{\alpha_1-1}{(x-u_1)}^{2a-2}{(1-2x+u_1)}^{\alpha_4-1} \, du_1
  \end{split}
\end{equation*}

Note that the first integral is well-defined and non-negative. On the other hand, the second integral is not defined: 
\begin{equation*}
  \begin{split}
    \int_{x/2}^{x} u_1^{\alpha_1-1}&{(x-u_1)}^{2a-2}{(1-2x+u_1)}^{\alpha_4-1} \, du_1 \\
    &\ge \int_{x/2}^x \min\left({\left(\frac{x}{2}\right)}^{\alpha_1-1}, x^{\alpha_1-1}\right){(x-u_1)}^{2a-2} \\ 
    &\hspace{3cm} \times \min\left({\left(1-\frac{3}{2}x\right)}^{\alpha_4-1}, {(1-x)}^{\alpha_4-1}\right) \, du_1 \\
    &= K(x) \int_{0}^{x/2} v^{2a-2} \, dv \\ 
    &= \begin{cases}
      \dfrac{K(x)}{2a-1} \lim_{t \to 0^+} \left[{(x/2)}^{2a-1} - t^{2a-1}\right] &\text{ if } a < 0.5 \\ 
      K(x) \lim_{t \to 0^+} \left[\log(x/2) - \log(t)\right] &\text{ if } a = 0.5
    \end{cases} \\
    &\to +\infty, 
  \end{split}
\end{equation*}
where $K(x)$ is a function of $x$. 

Based on this divergence, we conclude that if $0 < \alpha_2 = \alpha_3 \le 0.5$ and $x = y < 0.5$, $f_{X, Y}(x,y)$ is not defined. 
Notice that if $x = y \ge 0.5$, divergence problems still happen, since the problems appear when $u_1$ approximates $x$. Similar calculations show that if $x + y = 1$ and $0 < \alpha_1 = \alpha_4 \le 0.5$, the density is also not defined.
More generally, $f_{X,Y}(x,y)$ is not defined if $\alpha_1 +
\alpha_4 \le 1$ and $x + y = 1$; $\alpha_2 + \alpha_3 \le 1$ and $x = y$.

\end{document}